\documentclass[%
 reprint,
 amsmath,amssymb,
 aps,
 pra, 
 longbibliography,
 floatfix
]{revtex4-2}

\usepackage{footmisc,multirow}
\usepackage{subfigure} 
\usepackage{amsmath,slashed}
\usepackage{upgreek}
\usepackage{graphicx}
\usepackage{color}
\usepackage{comment}

\begin{document}

\title{Self-induced transparency and optical transients in atomic vapors}

\author{B. S. Cartwright}  
\affiliation{Department of Physics, Durham University, Durham DH1 3LE, United Kingdom}

\author{S. A. Wrathmall}
\email{s.a.wrathmall@durham.ac.uk} 
\affiliation{Department of Physics, Durham University, Durham DH1 3LE, United Kingdom}

\author{R. M. Potvliege}
\email{r.m.potvliege@durham.ac.uk}
\affiliation{Department of Physics, Durham University, Durham DH1 3LE, United Kingdom}


\begin{abstract}
The rapid turn-on of a strong, resonant, continuous wave laser field may trigger the formation of a transient oscillation akin to a train of damped solitons, before the vapor-field system relaxes into a stationary state. We study this transient dynamic on theoretical models of a rubidium vapor. We also consider doubly resonant V-systems, for which the transients take the form of trains of damped simultons. We compute the propagating field(s) by solving the Maxwell-Bloch equations, taking homogeneous broadening, Doppler broadening and the full hyperfine structure of the atoms into account. We also compare the actual fields to the stationary dnoidal fields predicted by the Maxwell-Bloch equations in conditions of self-induced transparency. A similar dynamics is expected to occur in any atomic vapor at the turn-on of a strong resonant continuous wave field provided the turn-on is sufficiently fast compared to relaxation.

\end{abstract}

\maketitle

\section{Introduction}

The interplay between optical fields and atomic vapors gives rise to a number of remarkable nonlinear phenomena.  
Some have attracted much attention in recent years, in particular electromagnetically induced transparency (EIT), dipole blockade and various co-operative effects. Others, such as self-induced transparency (SIT), do not figure as prominently in current research on atomic vapors. SIT  is a strong field phenomenon quenched by homogeneous broadening \cite{McCall,Crisp1969,Eberly1969,Lamb1971,Bolshov1985,Maimistov1990,seeAllen1975}. It is rarely relevant for the propagation of laser fields through thermal vapors such as those used in quantum technology applications, as these fields are typically too weak or applied for too long for this nonlinearity to play a role. As a consequence, much of the current experimental work in this area does not require consideration of SIT. However, beams controllable on a time scale of the order of the nanosecond and achieving GHz Rabi frequencies are sometimes used, which are parameters for which SIT may be relevant. The significance of this effect in atomic vapors has been considered in recent investigations, and prospects for harnessing this additional nonlinearity have been outlined \cite{Ogden2019,Arkhipov2020,Bai2020}. SIT also occurs in a variety of other media \cite{othermedia}, and SIT-related effects in the propagation of femtosecond or attosecond pulses have also been studied \cite{femto}.

SIT stems from the intrinsically nonlinear response of any atomic system to an applied electromagnetic field. Its best known manifestation is the formation of solitons in inhomogeneously broadened resonant two-level media: as long as spontaneous decay and other sources of homogeneous broadening can be neglected, a short and sufficiently strong laser pulse can propagate through a medium over arbitrarily long distances in the form of one or several shape-conserving hyperbolic secant pulses \cite{McCall}. The medium absorbs energy from the field on the raising edge of these pulses and entirely returns this energy on the trailing edge. These solitons thus propagate without damping.  They are not destroyed by inhomogeneous broadening because this cycle occurs in the same way for all the atoms or molecules forming the medium. Homogeneous broadening is deleterious to the effect, though, which in practice limits the range and lifetime of these solitons \cite{McCall,Ogden2019,Alhasan1992,Miklaszewski1994}.

SIT solitons were discovered and first studied for single-color fields. However, it was later noted that doubly resonant V-systems may also support twin pairs of solitons, called simultons, whereby two different fields resonant on different transitions co-propagate in the form of solitons with identical trajectories \cite{Kono81,seealso,Huang}.
Mathematically, these solitons and simultons are related to more general analytical solutions of the Maxwell-Bloch equations in which the field is described by a dnoidal Jacobian elliptic function $\mbox{dn}(\eta,k)$, where $\eta$ is a variable depending both on time and on  position and $k$ is the elliptic modulus of the function \cite{Crisp1969,Eberly1969,Hioe1994}.
Except in the limiting case of an elliptic modulus equal to 1, where they reduce to a single soliton, these solutions take the form of infinitely long pulse trains.

Such dnoidal solutions are particularly relevant to the present work, which is a numerical study of the nonlinear transients, akin to a train of soliton-like pulses, which develop when a strong continuous wave (CW) field is suddenly applied to an atomic vapor \cite{Crisp1972,Segard1990,deLamare1993}. While some of the details of our results are model specific, these nonlinear transients are generic. However, they may be so suppressed by decoherence as to be irrelevant if the applied field is too weak or is not turned on fast enough.

Although they may arise in a broad variety of systems, at least in some form, these nonlinear transients have not been as extensively studied as SIT solitons, dnoidal waves and simulton-like twin pulses.
The response of a medium to a resonant step pulse has been studied primarily in respect of the fast transients developing immediately after the incident pulse is applied, in advance of the main field (the Sommerfeld and Brillouin precursors and $0\pi$-pulses) \cite{Segard1990,Crisp1970,Rothenberg1984,Avenel1984,Pessina1991,LeFew2009,Wei2009,Macke2010,precursorpapers}. The delay in the arrival of a field propagating under self-induced transparency has been analysed in detail \cite{Crisp1972,Segard1990,Macke2010,Horovitz1982}. It is also known that the reshaping of a square pulse into a train of soliton-like pulses is an exactly solvable problem within the inverse scattering transform approach \cite{Kaup1977}. To the best of our knowledge, however, these soliton-like transients have been systematically studied only for single-color fields and, with the exception of Ref.~\cite{Macke2010}, only within calculations neglecting homogeneous broadening.
The present article addresses the role of the latter both for single-color fields and for two-color fields.

Specifically, we concentrate on the case of a thermal vapor interacting with a resonant optical field turned on from zero to a constant intensity on a nanosecond scale, such as a CW beam switched on by a fast Pockels cell. The rise time of the applied field is thus long compared to an optical period. However, it is short compared to the longitudinal relaxation time of the medium. We assume that this field, although strong, is nonetheless amenable to a description neglecting far off-resonance transitions and based on the rotating wave approximation. Besides 2-level systems interacting with a single field, we also consider V-systems of three or more states addressed by two different fields.
We do not consider
the cases of $\Lambda$ or ladder systems, as SIT in those systems depends sensitively on the initial populations and coherences, which are experiment-specific.
We stress that the transients studied in the present work originate from an applied field whose phase and intensity are constant once it has been turned on. Similar trains of soliton-like pulses may also develop under propagation of a periodically-pulsed applied field \cite{variapulses}.

Our theoretical approach is standard in this context. It is outlined in Sec.~\ref{section:theory} and in more detail in Appendix~\ref{appendix:Vdnoidal} (this appendix focuses on the case of a pair of fields interacting with three-level systems in a V-configuration and extends previous work on the topic). Numerical results illustrating the occurrence of transient pulse trains in 2-level systems and in V-systems are presented and discussed Secs.~\ref{section:results} and \ref{section:V}. Further results can be found in Ref.~\cite{Cartwright2022}.  The main body of the paper ends with brief conclusions, in Sec.~\ref{section:conclusions}.

\section{Mathematical framework}
\label{section:theory}

\subsection{SIT in two-state systems}
\label{section:twostate}
 
We start by concentrating on the simple case of a 2-state system consisting of a ground state (state~1) and an excited state (state~2)  coupled to each other by a linearly polarized field resonant on this transition. V-systems are considered in Sec.~\ref{section:V}.
We treat the field classically and write its electric field vector,
${\bf E}$, as the product of a slowly-varying envelope and a plane-wave carrier. For simplicity, we neglect transverse effects including diffraction and self-focusing and make the plane wave approximation (we briefly comment on the relevance of the transverse inhomogeneity of the field in Sec.~\ref{section:conclusions}). I.e., taking the unit polarization vector $\hat{\text{\boldmath{$\epsilon$}}}$ to be real and the field to propagate in the $z$-direction, we set
\begin{align}
{\bf E}_{}(z,t) &= \frac{1}{2}\, \hat{\text{\boldmath{$\epsilon$}}}_{}\, {\cal E}_{}(z,t)
\exp[-i\omega_{}(t - z/c)] + \mbox{c.c.}.
\end{align}
The electric field amplitude ${\cal E}_{}({z},t)$ may be complex.
In general, the spatial and temporal variation of ${\bf E}_{}(z,t)$ is governed by the equation
\begin{equation}
\frac{\partial^2 {\bf E}}{\partial z^2} - \frac{1}{c^2}
\frac{\partial^2 {\bf E}}{\partial t^2} = \mu_0 \,\frac{\partial^2 {\bf P}}{\partial t^2},
\end{equation}
where ${\bf P}$ is the medium polarization and $\mu_0$ is the vacuum permeability. As commonly done in this context,
we simplify this equation 
by making the ansatz 
\begin{align}
{\bf P}(z,t) = & \frac{1}{2}\,\hat{\text{\boldmath{$\epsilon$}}}_{} \, {\cal P}_{}({z},t)
\exp[-i\omega_{}(t - z/c)] + \mbox{c.c.}
\end{align}
and taking into account the fact that the complex amplitudes ${\cal E}_{}(z,t)$ and ${\cal P}_{}(z,t)$ vary slowly compared to the carrier. This yields 
the wave equation \cite{Lambropoulos2007}
\begin{align}
\frac{\partial {\cal E}_{}}{\partial z} + \frac{1}{c}
\frac{\partial {\cal E}_{}}{\partial t} &= \frac{i\omega}{2\epsilon_0 c}\,{\cal P}_{}(z,t). \label{eq:propap}
\end{align}
As the polarization depends on the electric field in a complicated way, this equation is non-linear. This dependence is not amenable to a treatment based on non-linear susceptibilities in the case of SIT.

We assume that the medium is homogeneous and composed of a single species of atoms, treat it quantum mechanically, and describe its state by the density operator $\hat{\rho}(z,t)$. For a thermal vapor,
\begin{equation}
    \hat{\rho}(z,t) = \int_{-\infty}^\infty\,f_{}(u_z)\hat{\rho}_v(z,t,u_z)\,{\rm d}u_z,
\label{eq:rhov}
\end{equation}
where $\hat{\rho}_v({z},t,u_z)$ is the density operator describing the state of the atoms moving with a velocity $u_z$ in the $z$-direction (the direction of propagation of the field), and $f_{}(u_z)$ is the Maxwell-Boltzmann distribution function:
\begin{equation}
f_{}(u_z) = \frac{1}{u_{\rm rms}\sqrt{\pi}}\exp(-u_z^2/u_{\rm rms}^2),
\end{equation}
where $u_{\rm rms}$ is the root-mean-square velocity of the atoms. 
For a vapor of temperature $T$ composed of atoms of mass $M$, $u_{\rm rms} = \sqrt{2k_{\rm B}T/M}$, where $k_{\rm B}$ is Boltzmann constant. Neglecting Doppler broadening amounts to replacing $f(u_z)$ by the delta function $\delta(u_z)$, in which case
\begin{equation}
    \hat{\rho}({z},t) \equiv \hat{\rho}_v({z},t,u_z = 0).
    \label{eq:noib}
\end{equation}

We assume that state 2 decays to state 1 at the rate $\Gamma_{12}$, that state 1 is stable, and that there is no other decoherence or dephasing mechanism. Accordingly, the $u_z$-dependent density operator $\hat{\rho}_v(z,t,u_z)$ evolves in time according to the Lindblad master equation
\begin{align}
\frac{\partial \hat{\rho}_v}{\partial t} &=
-\frac{i}{\hbar}\,[\hat{H},\hat{\rho_v}\,]
\nonumber \\ & \qquad + 
\frac{1}{2} \left(
2\, \hat{C} \hat{\rho_v} \, \hat{C}^\dagger - \hat{C}^\dagger \hat{C} \hat{\rho_v} - \hat{\rho_v}\, \hat{C}^\dagger \hat{C}
\right),
\label{eq:Lindblad}
\end{align}
where $\hat{C} = \sqrt{\Gamma_{12}}\, |1\rangle\langle 2|$.
This collapse operator is absent in the calculations neglecting homogeneous broadening, for which Eq.~(\ref{eq:Lindblad}) reduces to
\begin{align}
\frac{\partial \hat{\rho_v}}{\partial t} &=
-\frac{i}{\hbar}\,[\hat{H},\hat{\rho_v}\,].
\label{eq:Lindbladnohb}
\end{align}
We treat the coupling of the atoms with the field within the electric dipole approximation, make the rotating wave approximation, and pass to slowly varying variables. For each velocity class, doing so reduces the Hamiltonian to
\begin{align}
&\hat{H}({z},t) = - \hbar \Delta_{}|2 \rangle \langle 2 | -(\hbar/2) \big[\Omega_{}(z,t)
|2\rangle\langle 1| + \mbox{h.c.}\,\big]
\label{eq:H}
\end{align}
with
\begin{equation}
    \Omega_{}(z,t) =
{\cal E}_{}(z,t)\,\langle\, 2\, |\,
\hat{\text{\boldmath{$\epsilon$}}}_{} \cdot \hat{\bf D} \,|\, 1\rangle/\hbar,
\label{eq:Omegap}
\end{equation}
where $\hat{\bf D}$ is the dipole operator. For simplicity we assume that the field is on resonance ($\Delta = 0$) for $u_z= 0$. However, the thermal velocity of the atoms gives rise to non-zero values of the detuning $\Delta$ in the present calculations: for a velocity $u_z$ in the $z$-direction,
\begin{equation}
    \Delta_{} = -\omega_{}u_z/c.
\end{equation}

We denote the coherence $\langle 2 | \hat{\rho}(z,t) | 1 \rangle$ by $\rho_{21}(z,t)$. In terms of this quantity,
\begin{align}
    {\cal P}_{}(z,t) &= 2 N_{\rm d} \, 
    \langle\, 1\, |\,
\hat{\text{\boldmath{$\epsilon$}}}_{} \cdot \hat{\bf D} \,|\, 2\, \rangle\, \rho_{21}(z,t),
\label{eq:polp}
\end{align}
where $N_{\rm d}$ is the number density of the atoms in the medium. 
Combining this last equations with Eqs.~(\ref{eq:propap}) and (\ref{eq:Omegap}) gives
\begin{align}
  \frac{\partial \Omega_{}}{\partial z} + \frac{1}{c}\,\frac{\partial \Omega_{}}{\partial t} &= i \mu_{} \,
    \rho_{21}(z,t), \label{eq:Omegappropa}
\end{align}
where $\mu_{}$ is the propagation coefficient:
\begin{align}
    \mu_{} &= \frac{\omega_{}N_{\rm d}}{\hbar \epsilon_0 c}\, |\langle 2 |\,
\hat{\text{\boldmath{$\epsilon$}}}_{} \cdot \hat{\bf D} \,|\, 1\, \rangle|^2.
\end{align}
The propagation of the field through the medium is governed by Eq.~(\ref{eq:Omegappropa}) together with Eq.~(\ref{eq:Lindblad}) or Eq.~(\ref{eq:Lindbladnohb}). It is worth noting that Eq.~(\ref{eq:Omegappropa}) admits real solutions in the present case because the coherence $\rho_{21}(z,t)$ is purely imaginary when $\Delta = 0$. 

As long as homogeneous broadening can be neglected, these propagation equations admit stationary solutions depending on $z$ and $t$ only through the combination $t - z/v$, where $v$ is a certain velocity. Of particular interest here are dnoidal wave solutions described by Jacobian elliptic functions --- i.e., solutions for which \cite{Crisp1969,Eberly1969,warning}
\begin{equation}
    {\cal E}_{}(z,t) = {\cal E}_{0}\,\mbox{dn}(\eta,k)
\end{equation}
and
\begin{equation}
    {\Omega}_{}(z,t) = {\Omega}_{0}\,\mbox{dn}(\eta,k)
    \label{eq:Omegacdn}
\end{equation}
for a constant elliptic modulus $k$ ($0 < k \leq 1$).
These solutions depend on $z$ and $t$ through the variable $\eta$ defined by the equation
\begin{equation}
    \eta = (t-z/v)/\tau
    \label{eq:etadefined}
\end{equation}
with
\begin{equation}
    \tau = 2 / \Omega_{0}.
    \label{eq:taudefined}
\end{equation}
${\cal E}_{{}0}$,
${\Omega}_{{}0}$ and the elliptic modulus $k$ of the $\mbox{dn}$ function are constants. Moreover 
\begin{equation}
\frac{1}{v} = \frac{1}{c} + \frac{\mu_{}\tau^2}{2k^2}
\label{eq:vdnoidal}
\end{equation}
if Doppler broadening is ignored, or
\begin{equation}
    \frac{1}{v} = \frac{1}{c} +
    \frac{\mu_{} \tau^2}{2} \int_{-\infty}^\infty
    \frac{g_{}(\Delta_{})\,{\rm d}\Delta_{}}{[(k^2 - \Delta_{}^2\tau^2)^2  + 4 \Delta_{}^2\tau^2]^{1/2}}
    \label{eq:vdnoidalib}
\end{equation}
if Doppler broadening is taken into account \cite{Crisp1969}. The function $g(\Delta_{})$ appearing in this last equation is the Maxwell-Boltzmann distribution expressed as a function of the detuning:
\begin{equation}
    g(\Delta_{}) \equiv (c/\omega_{})f(c \Delta_{}/\omega_{}).
\label{eq:g}
\end{equation}
In terms of the corresponding inhomogeneous transverse relaxation time $T_2^* = c \sqrt{2}/\omega u_{\rm rms}$,
\begin{equation}
    g(\Delta) = \frac{T_2^*}{\sqrt{2\pi}}\,\exp[-(T_2^*\Delta)^2/2].
\end{equation}

For $0 < k < 1$, this dnoidal wave solution takes the form of a stationary and infinitely long train of identical shape-conserving pulses repeating themselves periodically and traveling through the medium at the $k$- and $\Omega_{{}0}$-dependent speed $v$.
The electric field amplitude ${\cal E}_{}(z,t)$
oscillates between $(1-k^2)^{1/2}{\cal E}_{{}0}$ and ${\cal E}_{{}0}$ through each pulse. The period of this oscillation (e.g., the time interval between two consecutive maxima) is $2\tau K(k)$, where $K(\cdot)$ is the complete elliptic integral of the first kind \cite{Kdefined}. 
This infinite pulse train morphs into a single pulse when $k = 1$ since $K(k) \rightarrow \infty$ when $k \rightarrow 1$ and
\begin{equation}
    \lim_{k \rightarrow 1}\,\mbox{dn}(\eta,k) \equiv \mbox{sech}(\eta).
\end{equation}
Eqs.~(\ref{eq:Omegacdn}), (\ref{eq:vdnoidal}) and (\ref{eq:vdnoidalib}) thus reduce to the well known relations describing an SIT soliton in this limit \cite{McCall}.

As seen from the above, Doppler broadening changes the speed at which the field propagates in the medium but not the functional form of $\Omega(z,t)$. The propagation speed is always faster than in the absence of Doppler broadening since
\begin{equation}
    [(k^2 - \Delta^2\tau^2)^2+4\Delta^2\tau^2]^{1/2} > k^2
\end{equation}
for any $\Delta\tau \not= 0$. Physically, the increase in $v$ can be understood as arising from an effective reduction in the density of the atoms interacting with the field.

We also note that Eq.~(\ref{eq:vdnoidalib}) can be simplified in the limit where the time parameter $\tau$ is much larger than the inhomogeneous time $T_2^*$.
The function $g(\Delta)$ indeed varies more slowly around $\Delta = 0$ than the denominator of Eq.~(\ref{eq:vdnoidalib}) in this case, and therefore
\begin{align}
    \frac{1}{v} &\approx \frac{1}{c} +
    \frac{\mu \tau^2}{2}\, g_{}(0)\int_{-\infty}^\infty
    \frac{{\rm d}\Delta}{[(k^2 - \Delta^2\tau^2)^2+4\Delta^2\tau^2]^{1/2}} \nonumber \\
    &\approx \frac{1}{c} + \frac{\mu \tau g(0)}{1 + k'}\,K\left(\frac{2\sqrt{k'}}{1+k'}\right),
\end{align}
where $k' = (1-k^2)^{1/2}$. The speed $v$ in general decreases when the temperature of the vapor increases because the product $N_{\rm d}$ normally increases faster with $T$ than $g(0)$ decreases. This speed also varies with the elliptic modulus $k$: $v$ increases with $k$ and is largest for $k=1$ (i.e., in the limit where the dnoidal wave reduces to a sech-soliton).

These dnoidal solutions (with $k\not= 1$) only
apply to cases where the field can be treated as being perfectly stationary and where homogeneous broadening is negligible. They are also unphysical in that they have no beginning, no end and an infinite spatial extension. These solutions are nonetheless useful as they provide some insight on the behavior of more realistic systems, as we discuss in Sec.~\ref{section:results}. 

\subsection{SIT in three-state V-systems}
\label{section:threestate}

SIT is not restricted to single fields. In particular, Konopnicki and Eberly have shown that doubly resonant three-level V-systems may support pairs of co-propagating sech pulses, called simultons \cite{Kono81,seealso}: each of the two fields propagates as a single soliton of the same width, speed and position, but not necessarily of the same amplitude, as for the other field.
It is not necessary for this to happen that both fields are sufficiently strong to propagate as solitons in the absence of the other field. A strong pulse resonant on one transition may thus transport a very weak field resonant on another transition over distances much exceeding those predicted by the Beer-Lambert law.
The former may thus control the propagation of the latter in this case, in a form of electromagnetically induced transparency in the broad sense of the term. The effect has been called ``soliton-induced transparency" \cite{Kozlov2009}. 
The underlying mathematical theory guarantees the stability of these simultons only for systems for which the two transitions have the same oscillator strength, are inhomogeneously broadened in the same way and relaxation can be ignored, which are conditions difficult to meet in experiments. However, these conditions are not stringent and can be relaxed without suppressing the co-propagation of twin pulses over significant distances \cite{Ogden2019,Bolshov1982b,preprint,Bolshov1988,Kozlov1998,Denisova1998,Kozlov1999,Paspalakis2000,KK2010,Fedotova2014}.

Simulton analytical solutions of the Maxwell-Bloch equations for such systems are limiting cases of more general solutions in which the two fields co-propagate as twin pairs of waves described by Jacobian elliptic functions \cite{Hioe1994}. 
The relevant mathematical details are given in Appendix~\ref{appendix:Vdnoidal} for the case of a 3-level V-system addressed by two fields. A first field, the ``probe field", couples states 1 and 2; a second field, the ``coupling field", couples states 1 and 3. Each of these fields is resonant on the corresponding transition. We denote the respective Rabi frequencies by ${\Omega}_{\rm p}(z,t)$ and ${\Omega}_{\rm c}(z,t)$.
In the ideal case where $\mu_{\rm p} = \mu_{\rm c}$, the distribution function $g(\Delta)$ is the same for the two fields and homogeneous broadening is negligible, this system can support twin dnoidal waves for which
\begin{align}
    &{\Omega}_{\rm p}(z,t) = 
    {\Omega}_{{\rm p}0}\,\mbox{dn}(\eta,k) \label{eq:25}
\end{align}
and
\begin{align}
    &{\Omega}_{\rm c}(z,t) = 
    {\Omega}_{{\rm c}0}\,\mbox{dn}(\eta,k), 
    \label{eq:26}
\end{align}
with $\eta$ and $v$ defined as above but with Eq.~(\ref{eq:taudefined}) replaced by
\begin{equation}
    \tau = 2 / (\Omega_{{\rm p}0}^2 + \Omega_{{\rm c}0}^2)^{1/2}.
\label{eq:tau2colour}
\end{equation}
The speed at which these twin waves propagate, $v$, is related to the temporal width $\tau$ in the same way as for a single-field dnoidal wave. However, this speed is the same for the two fields, here, and depends on the Rabi frequency amplitudes $\Omega_{{\rm p}0}$ and $\Omega_{{\rm c}0}$ only through the combined Rabi frequency amplitude $(\Omega_{{\rm p}0}^2 + \Omega_{{\rm c}0}^2)^{1/2}$. Soliton-induced transparency thus generalizes into a ``dnoidal-wave-induced transparency" whereby a strong coupling field carries a weak probe field through the medium as a co-propagating twin dnoidal wave.

We show, in Appendix~\ref{appendix:Vdnoidal}, that these analytical solutions can be extended to the case where the two transitions have different oscillator strengths if one of the two fields is much weaker than the other. The stronger field is still described by Eq.~(\ref{eq:Omegacdn}) whereas the weaker one has the more complicated functional form given by Eq.~(\ref{eq:Omegapweak3}).
These analytical solutions only
apply to systems for which all broadening mechanisms are negligible and the two fields can be treated as being perfectly stationary. However, and as we show in Sec.~\ref{section:V}, switching on a doubly resonant CW field may create transient pulse trains akin to a twin pair of dnoidal waves even when these conditions are not met.

\section{Non-linear transients}
\label{section:results}

We first consider the case of a 2-state medium exposed to an incident field smoothly turned on from 0 to a constant intensity $I^{\rm in}$. The medium extends in the $z$-direction from $z=0$ (the point at which the incident field is applied) to arbitrarily larges values of $z$. Given ${\cal E}(z,t)$ at $z=0$, we calculate the field inside the medium by numerically integrating Eq.~(\ref{eq:Omegappropa}) and either Eq.~(\ref{eq:Lindblad}) or Eq.~(\ref{eq:Lindbladnohb}) \cite{Ogden2019,Potvliege2025}. The numerical methods we used to this effect are described in detail in Ref.~\cite{Potvliege2025}.

The results presented in this section are calculated for a linearly polarized 780~nm laser field propagating in an isotopically pure ${}^{85}$Rb vapor. The field is resonant on the D$_2$ transition. We assume a temperature of 80$^\circ$C. Accordingly, we set $N_{\rm d} = 1.55 \times 10^{18}$~m$^{-3}$ and $u_{\rm rms} = 263$~m~s$^{-1}$, $\Gamma_{12} = 2\pi \times 6.067$~MHz and
\begin{equation}
\langle\, 1\, |\,
\hat{\text{\boldmath{$\epsilon$}}}_{} \cdot \hat{\bf D} \,|\, 2\rangle = 
\langle 5 {\rm S}_{1/2} || \, er \, || 5 {\rm P}_{3/2} \rangle/\sqrt{3} 
\end{equation}
with $\langle 5 {\rm S}_{1/2} || \, er \, || 5 {\rm P}_{3/2} \rangle = 3.584\times 10^{-29}$~C~m \cite{Steck}. 
The atoms forming the vapor are initially in their ground state. We also assume that 
\begin{equation}
    {\cal E}_{}(z=0,t) = {\cal E}_{}^{\rm in}\, F(t), 
\label{eq:Eoft}
\end{equation}
where ${\cal E}_{}^{\rm in}$ is a constant and
\begin{equation}
    F(t) = \begin{cases}
    0 & t < \mbox{$0$}, \\
    \cos^2[\pi (\mbox{$2$~ns}-t)/\mbox{4~ns}] & 0 \leq t \leq \mbox{$2$~ns} \\
    1 & t > \mbox{$2$~ns}.
    \end{cases}
\label{eq:Foft}
\end{equation}
The applied field is thus smoothly turned on over 2~ns, between $t = 0$ and $t=2$~ns, after which it is stationary and has the constant intensity
\begin{equation}
I^{\rm in} = \epsilon_0 c\,|{\cal E}^{\rm in}|^2/2.
\end{equation}
While slow compared to an optical period, this 2-ns turn-on is rapid compared to the longitudinal and intrinsic transverse relaxation times of the medium, $T_1 = 1/\Gamma_{12} = 26$~ns and $T_2 = 2 T_1$, respectively. We focus on the case of fields strong enough for driving Rabi oscillations whose period is short compared to $T_1$ and $T_2$.

\begin{figure}[t]
    \centering
\includegraphics[width=0.48\textwidth]{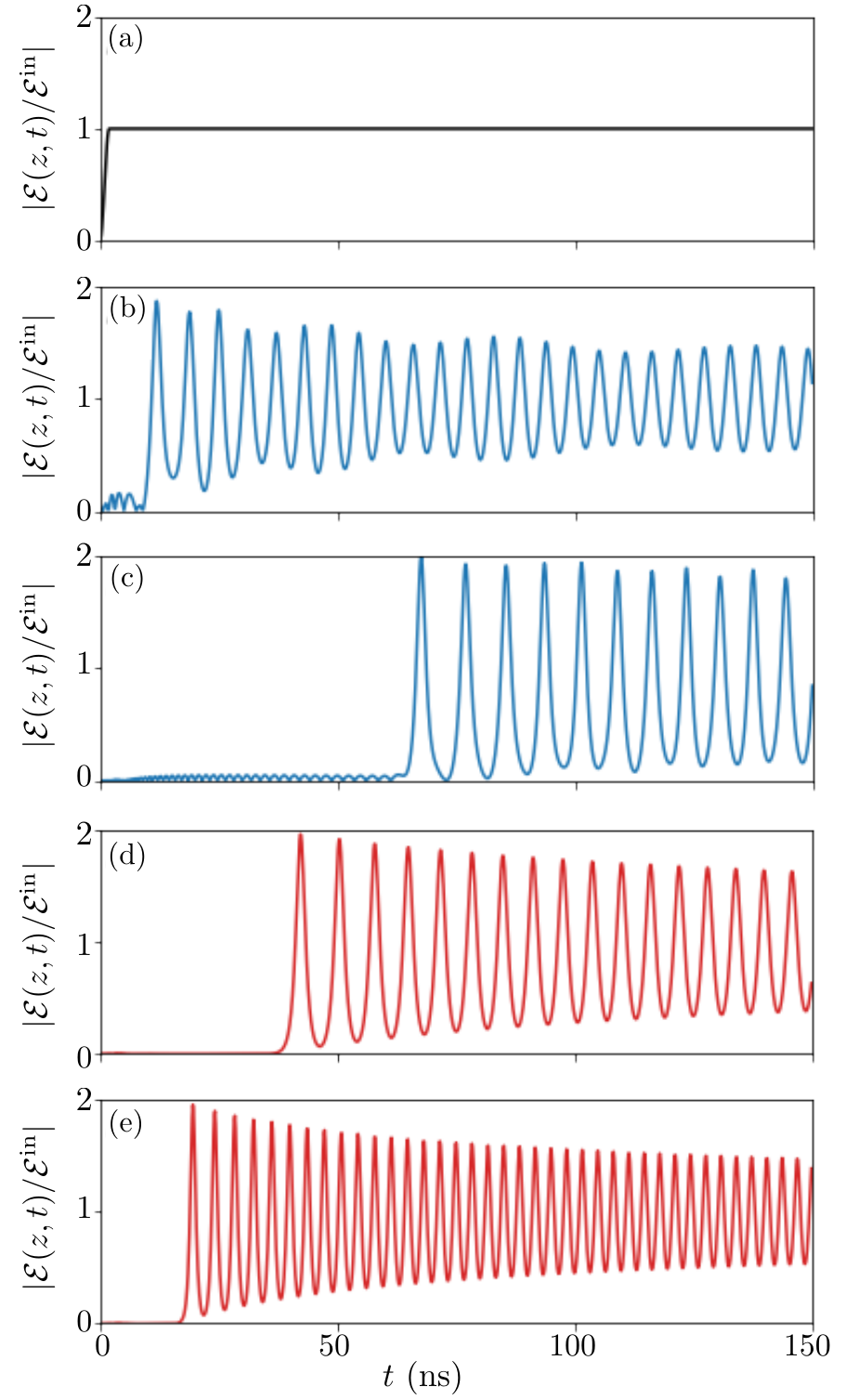}
\caption{
(a) The applied field in the 2-state model discussed in Sec.~\ref{section:results}. (b) The field at $z=3$~mm as obtained when homogeneous broadening and Doppler broadening are both neglected and $I^{\rm in}= 5$~W~cm$^{-2}$. (c) The same as (b) but for a propagation distance of 30~mm.  (d) The same as (c) but now with Doppler broadening taken into account. (e) The same as (d) but for $I^{\rm in}= 15$~W~cm$^{-2}$.
}
\label{fig:1}
\end{figure}
\begin{figure}[t]
    \centering
    \includegraphics[width=0.48\textwidth]{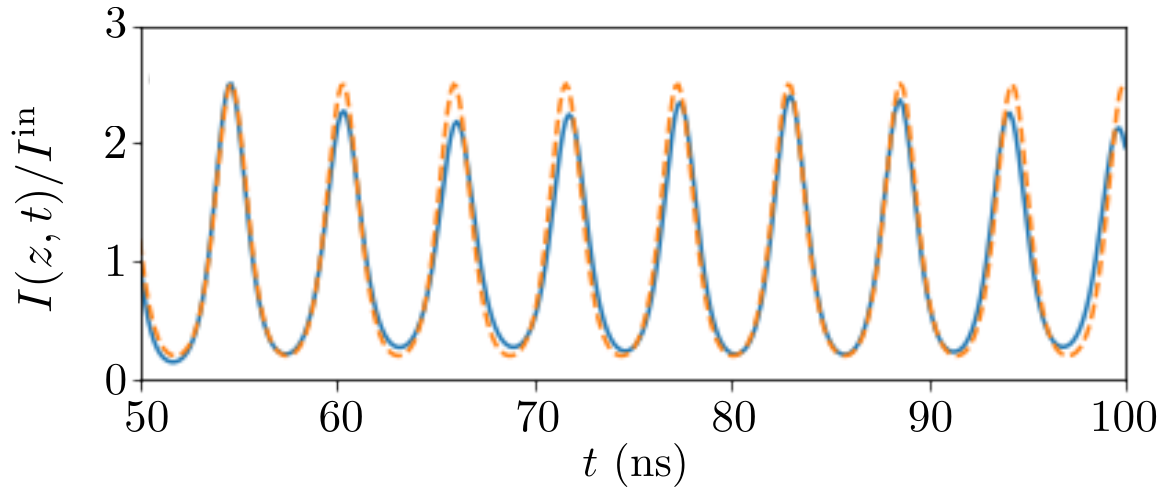}
\caption{Solid curve: The intensity of the field at a distance of $3$~mm inside the medium for $I^{\rm in} = 5$~W~cm$^{-2}$, as obtained when homogeneous broadening is neglected (Doppler broadening is included in the calculation). Dashed curve: the dnoidal function best fitting the solid curve at $t \approx 55$~ns.}
\label{fig:2}
\end{figure}
\begin{figure}[t]
    \centering
    \includegraphics[width=0.48\textwidth]{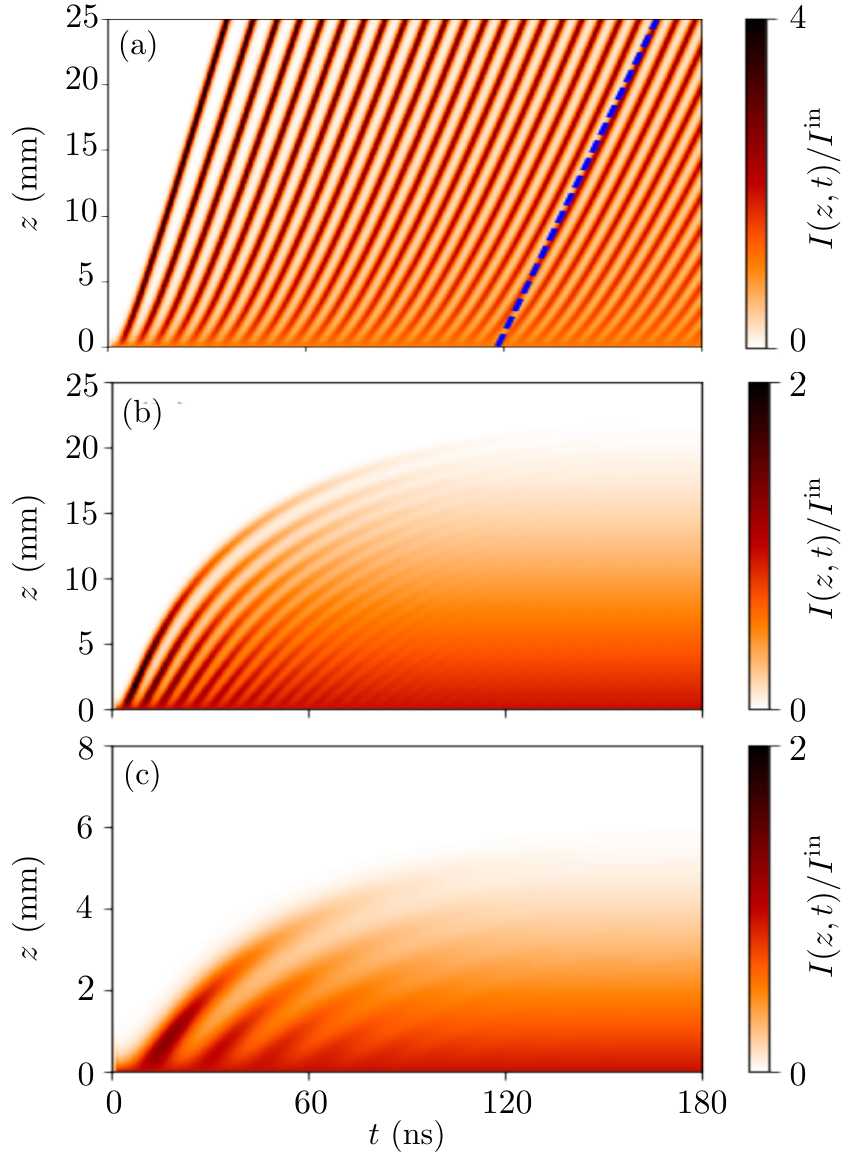}
\caption{The variation of the field in the 2-state model discussed in Sec.~\ref{section:results}. Doppler broadening is included in the calculation. (a) The field for $I^{\rm in} = 5$~W~cm$^{-2}$ when homogeneous broadening is neglected. The dashed line indicates the velocity of the dnoidal wave that most closely matches the field between $t = 120$~ns and $t = 180$~ns at $z = 25$~mm. (b) The same as (a) but here with homogeneous broadening taken into account. (c) The same as (b) for $I^{\rm in} = 500$~mW~cm$^{-2}$.
\label{fig:3}
}
\end{figure}

The time of arrival of the transmitted field as a function of depth in the medium for such 2-state systems has been studied both theoretically \cite{Crisp1972,deLamare1993,Horovitz1982,Pessina1991,Macke2010} and experimentally \cite{Segard1990,deLamare1993}. These earlier publications, particularly Ref.~\cite{Horovitz1982}, also noted and commented on the oscillatory character of the envelope of the transmitted field, although not always in much detail. Its variation for the system described above is illustrated by Fig.~\ref{fig:1}. (In this figure as in the rest of the paper, the time represented on the horizontal axis is the absolute time, $t$, rather than the retarded time, $t' = t - z/c$.)
Part~(a) of the figure shows the electric field amplitude of the applied field at the front of the medium. Parts (b) and (c) show the transmitted field after propagation over a distance of  3~mm or 30~mm respectively, 
as obtained when all broadening mechanisms are neglected. A value of 5~W~cm$^{-2}$ is assumed for $I^{\rm in}$, which is far larger than the saturation intensity for this transition, 2.5~mW~cm$^{-2}$ \cite{Steck}.  At first ${\cal E}(z,t)$ oscillates non-harmonically about zero, forming a train of low intensity $0\pi$ pulses \cite{Crisp1970,Rothenberg1984,Grieneisen1972,Hamadani1974,Matusovsky1996b}.
In this initial stage, the transmitted intensity is roughly proportional to $I^{\rm in}$. These initial transients are then followed by the arrival of the main field, whose electric field amplitude also oscillates. This oscillation continues indefinitely here because decoherence is neglected. As shown by parts (d) and (e) of Fig.~\ref{fig:1}, by Fig.~\ref{fig:2} and by Fig.~\ref{fig:3}, including Doppler broadening in the calculation suppresses the initial $0\pi$ pulses but not the oscillations of the main field.

As was already recognized in the early 1970s~\cite{Crisp1972}, the oscillation of the electric field amplitude of the main field for such systems is similar to that of a dnoidal solution of the Maxwell-Bloch equation, i.e., a solution of the form ${\cal E}(z,t) = {\cal E}_0\,\mbox{dn}(\eta,k)$ where $\eta$ is defined by Eq.~(\ref{eq:etadefined}). For example, we show, in Fig.~\ref{fig:2}, that ${\cal E}(z,t)$ can be closely approximated by a dnoidal function over a sufficiently small range of values of $t$ even when Doppler broadening is included in the calculation. The dnoidal wave model does not capture all the features of the solution for a finite step pulse, though: contrary to this model, the actual field is not infinite in spatial extension, is zero before the incident pulse is applied to the medium, and does not have a constant group velocity. The latter is related to the amplitude ${\cal E}_0$ and the elliptic modulus $k$ by Eqs.~(\ref{eq:etadefined})--(\ref{eq:vdnoidalib}), and for the analytical solution of best fit these two parameters both vary with $z$ and $t$. In particular, $k$ increases with $z$: as it travels through the medium, the incident step pulse morphs into a train of isolated solitons more and more spaced in time --- compare, e.g., parts (b) and (c) of the figure. The effect is clearly visible in Fig.~\ref{fig:3}(a). The time separation between the maxima of ${\cal E}(z,t)$ does not remain constant as the field propagates. Our results disagree with the conclusions of Ref.~\cite{Horovitz1982} on this point.

The relation between the oscillation of the main field and the time variation of its area,
\begin{equation}
    \theta(z,t) = \int_{-\infty}^t \Omega(z,t')\,{\rm d}t',
\end{equation}
is analysed in Ref.~\cite{Cartwright2022}.
Numerical calculations show that each of the pulses forming the main field is in one-to-one correspondence with an increase of $\theta(z,t)$ by $2\pi$, as in the case of SIT solitons. The rate of formation of these pulses therefore increases with $I^{\rm in}$, as is readily noticed by comparing Fig.~\ref{fig:1}(d) to Fig.~\ref{fig:1}(e).

While the analytical theory does not capture all the features of the numerical solution, it is nonetheless consistent with the latter in predicting (1) that inhomogeneous broadening increases the group velocity of the main field and (2) that the group velocity  increases and the width of the pulses decreases when the intensity increases --- e.g., compare  Fig.~\ref{fig:1}(c) to Fig.~\ref{fig:1}(d) and Fig.~\ref{fig:1}(d) to Fig.~\ref{fig:1}(e). The analytical theory is also consistent with the numerical results in correctly predicting the group velocity, as shown by the close agreement of the dashed blue line with the numerical solution in Fig.~3(a) [this dashed line represents the trajectory a maximum of the field would follow if its velocity was given by Eq.~(\ref{eq:vdnoidalib})]. It is worth noting that the field propagates in the medium at a speed much smaller than $c$ in all these results.

Comparing parts~(a) and (b) of Fig.~\ref{fig:3} shows how spontaneous decay affects the propagation dynamics. The parameters of the system are the same in these two sets of results, but spontaneous decay is taken into account in part~(b) and neglected in part~(a). One can observe that the calculated field does no longer keep oscillating indefinitely when spontaneous decay is included. The conspicuous oscillation found in the absence of homogeneous broadening instead reduces to a short train of transients. Also, the group velocity of the field now decreases rather than increases as the field propagates. As time progresses, the oscillation of ${\cal E}(z,t)$ disappears, the field tends to a stationary state, and the propagation regime changes from self-induced transparency to strong field absorption.
As expected from the above, reducing $I^{\rm in}$ leads to fewer and broader transients and to a more rapid absorption of the field [Fig.~\ref{fig:3}(c)].

Similar transients may also occur in experiments using combinations of fields of different frequencies, as shown in the next section.

\section{Transients in V-systems}
\label{section:V}

\begin{figure}[t]
    \centering
    \includegraphics[width=0.48\textwidth]{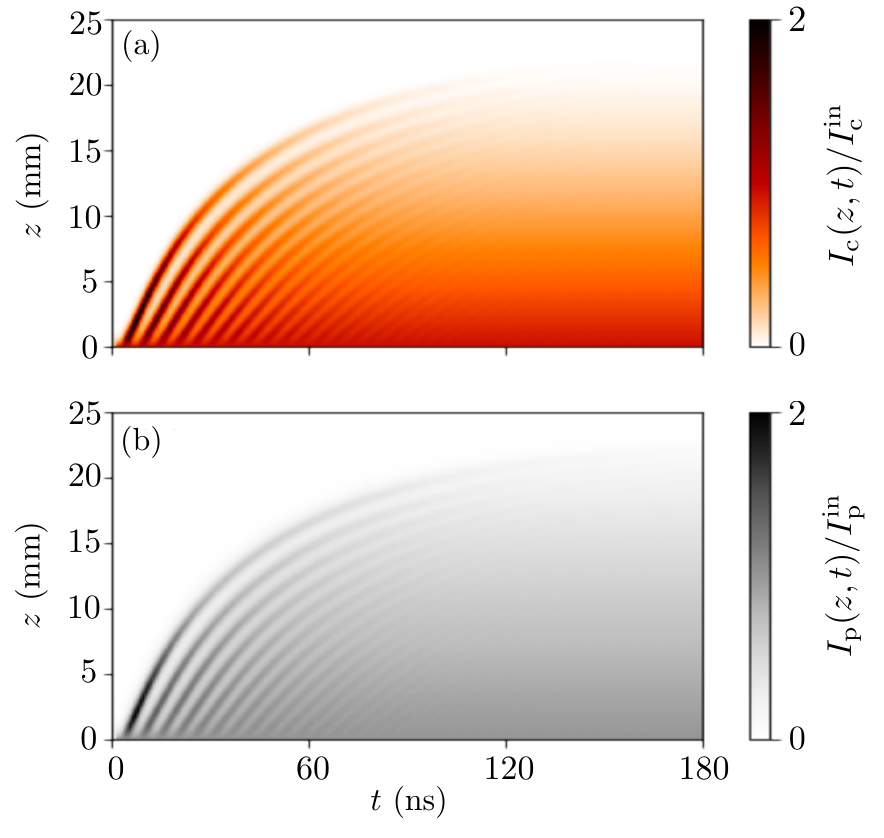}
\caption{
(a) Coupling field with $I_{\rm c}^{\rm in} = 5$~W~cm$^{-2}$. (b) Probe field with $I_{\rm p}^{\rm in} = 1$~$\mu$W~cm$^{-2}$ when co-propagating with the coupling field shown in (a).
}
\label{fig:4}
\end{figure}

\begin{figure}[t]
    \centering
    \includegraphics[width=0.48\textwidth]{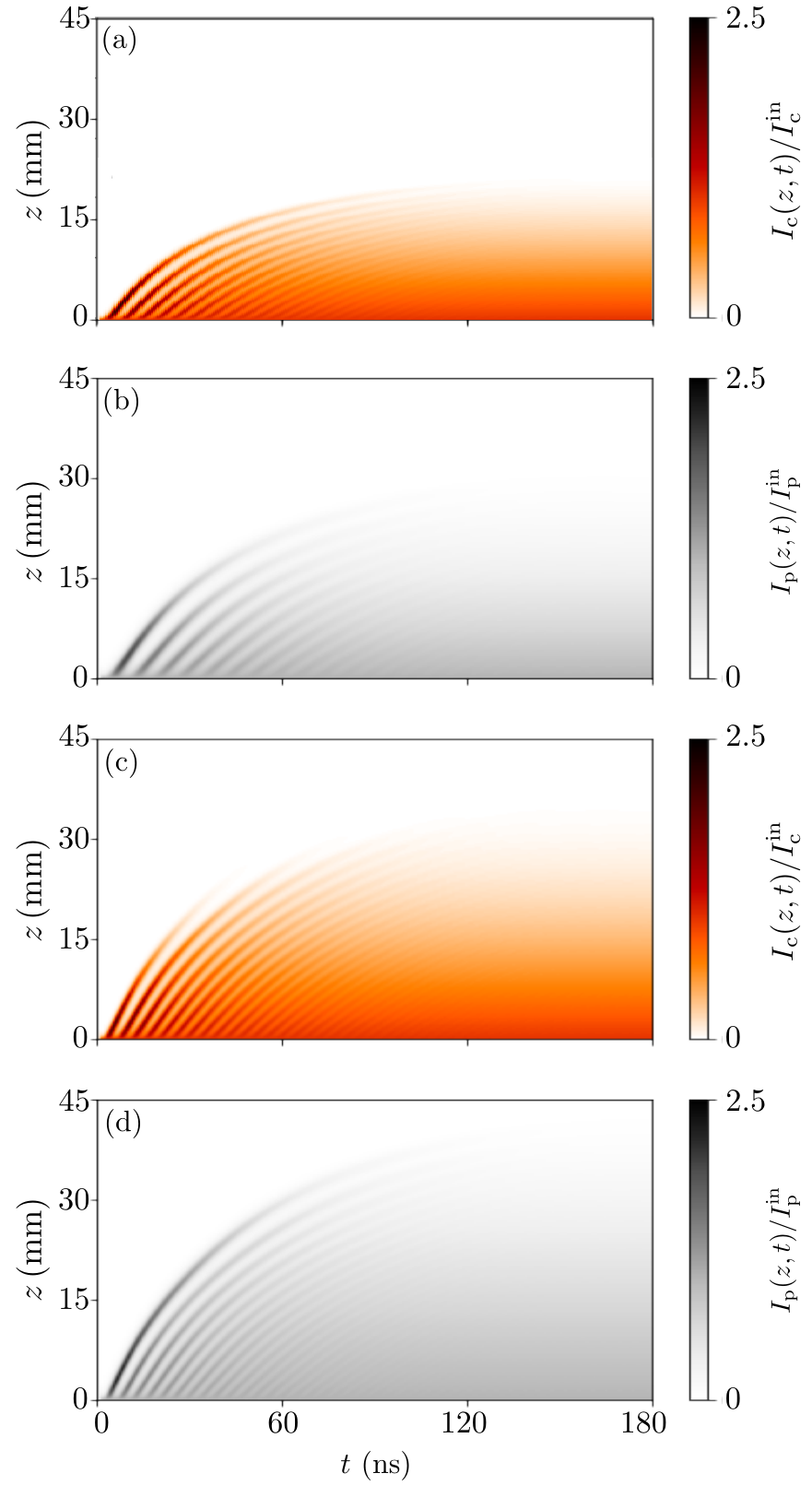}
    \caption{(a) Coupling field when  propagating in absence of the probe field for $I_{\rm c}^{\rm in} = 5$~W~cm$^{-2}$. (b) Probe field when  propagating in absence of the coupling field for $I_{\rm p}^{\rm in} = 5$~W~cm$^{-2}$. (c) Coupling field when co-propagating with the probe field for $I_{\rm c}^{\rm in} = I_{\rm p}^{\rm in} = 5$~W~cm$^{-2}$. (d) Probe field when co-propagating with the coupling field for $I_{\rm c}^{\rm in} = I_{\rm p}^{\rm in} = 5$~W~cm$^{-2}$.}
\label{fig:5}
\end{figure}

These calculations are extended to multistate systems addressed by a pair of step pulses in this section. As in Sec.~\ref{section:results}, we assume that these two fields propagate in an isotopically pure $^{85}$Rb vapor at a temperature of 80$^\circ$C. To start, we describe the atoms forming the medium as 3-state systems consisting of a single ground state (state~1), a single $5{\rm P}_{1/2}$ state (state~2 here) and a single $5{\rm P}_{3/2}$ state (state~3 here). One of the fields (the probe field) is resonant on the D$_1$ transition, the other (the coupling field) is resonant on the D$_2$ transition. Both are linearly polarized. In the notation of Appendix~\ref{appendix:Vdnoidal},
\begin{equation}
\langle\, 1\, |\,
\hat{\text{\boldmath{$\epsilon$}}}_{{\rm p}} \cdot \hat{\bf D} \,|\, 2\rangle = 
\langle 5 {\rm S}_{1/2} || \, er \, || 5 {\rm P}_{1/2} \rangle/\sqrt{3} 
\end{equation}
with $\langle 5 {\rm S}_{1/2} || \, er \, || 5 {\rm P}_{1/2} \rangle = 2.538\times 10^{-29}$~C~m and 
\begin{equation}
\langle\, 1\, |\,
\hat{\text{\boldmath{$\epsilon$}}}_{\rm c} \cdot \hat{\bf D} \,|\, 3\rangle = 
\langle 5 {\rm S}_{1/2} || \, er \, || 5 {\rm P}_{3/2} \rangle/\sqrt{3} 
\end{equation}
with $\langle 5 {\rm S}_{1/2} || \, er \, || 5 {\rm P}_{3/2} \rangle = 3.584\times 10^{-29}$~C~m \cite{Steck}.
The corresponding propagation coefficients differ significantly: 
$\mu_{\rm p}/\mu_{\rm c} = 0.49$ here.
The spontaneous decay rates, $\Gamma_{12}$ and $\Gamma_{13}$, are $2\pi \times 5.750$~MHz and $2\pi \times 6.067$~MHz, respectively \cite{Steck}. We assume that the applied fields are simultaneously turned on to constant intensities $I_{\rm p}^{\rm in}$ and $I_{\rm c}^{\rm in}$ as per Eqs.~(\ref{eq:Eoft}) and (\ref{eq:Foft}).
Figs.~\ref{fig:4} and \ref{fig:5} illustrate how these two fields propagate through the medium in this 3-state model. The calculation includes both homogeneous broadening and Doppler broadening.

The probe field is considerably weaker than the coupling field in the results shown in Fig.~\ref{fig:4}
(these results were calculated for $I_{\rm p}^{\rm in} = 1$~$\mu$W~cm$^{-2}$ and $I_{\rm c}^{\rm in} = 5$~W~cm$^{-2}$).  The probe field does not significantly affect the coupling field here, and the latter propagates as if the former was not present. In agreement with Sec.~\ref{section:results}, the coupling field initially undergoes a transient oscillation before settling into a CW steady state [Fig.~\ref{fig:4}(a)]. By contrast, the weak probe field would be almost completely absorbed over the first few mm of propagation in the medium if the coupling field was not present \cite{explainalpha}. However, and as seen from Fig.~\ref{fig:4}(b), the probe field travels significantly farther when the latter is present and undergoes the same transient oscillation. The strong coupling field effectively transports the weak probe field. Qualitatively, the effect is similar to the ``dnoidal-wave-induced transparency" discussed in Appendix \ref{appendix:unequal}, although the mathematical theory outlined in the appendix assumes no homogeneous or Doppler broadening. 
Since this theory also assumes that the probe field is very weak, it does not apply either when both fields are strong even when broadening is negligible.

How the two fields co-propagate when both are strong is illustrated by the results presented in Fig.~\ref{fig:5} ($I_{\rm p}^{\rm in} =  I_{\rm c}^{\rm in} = 5$~W~cm$^{-2}$ for these results). Parts (a) and (b) of the figure show the coupling and probe fields, respectively, when propagating in the absence of the other field. While both fields exhibit transient oscillations and penetrate far inside the medium at these high values of $I_{\rm p}^{\rm in}$ and $I_{\rm c}^{\rm in}$, the transients oscillate with a slower period for the probe field (the slower period of oscillation is consistent with the slower increase of the incident pulse area for this field, due to the smaller transition dipole moment). However, both fields travel significantly farther when co-propagating together, and their transients oscillate with the same period and phase [Fig.~\ref{fig:5}(c,d)].

These transients and SIT in general differ from EIT in the usual sense of the term in being time-dependent and not arising from a destructive interference between excitation pathways. The transparency observed in the steady-state regime which develops as these transients subside is often referred to as EIT, although EIT in V-systems \cite{Higgins2021} arises primarily from Autler-Townes splitting: a sufficiently strong coupling field splits the energies of the dressed states of the system out of resonance with a weak probe field, and the latter can then propagate through the medium as though it were transparent \cite{AbiSalloum2010,Khan2016,Zhu2013}.

\begin{figure}[t]
    \centering
    \includegraphics[width=0.48\textwidth]{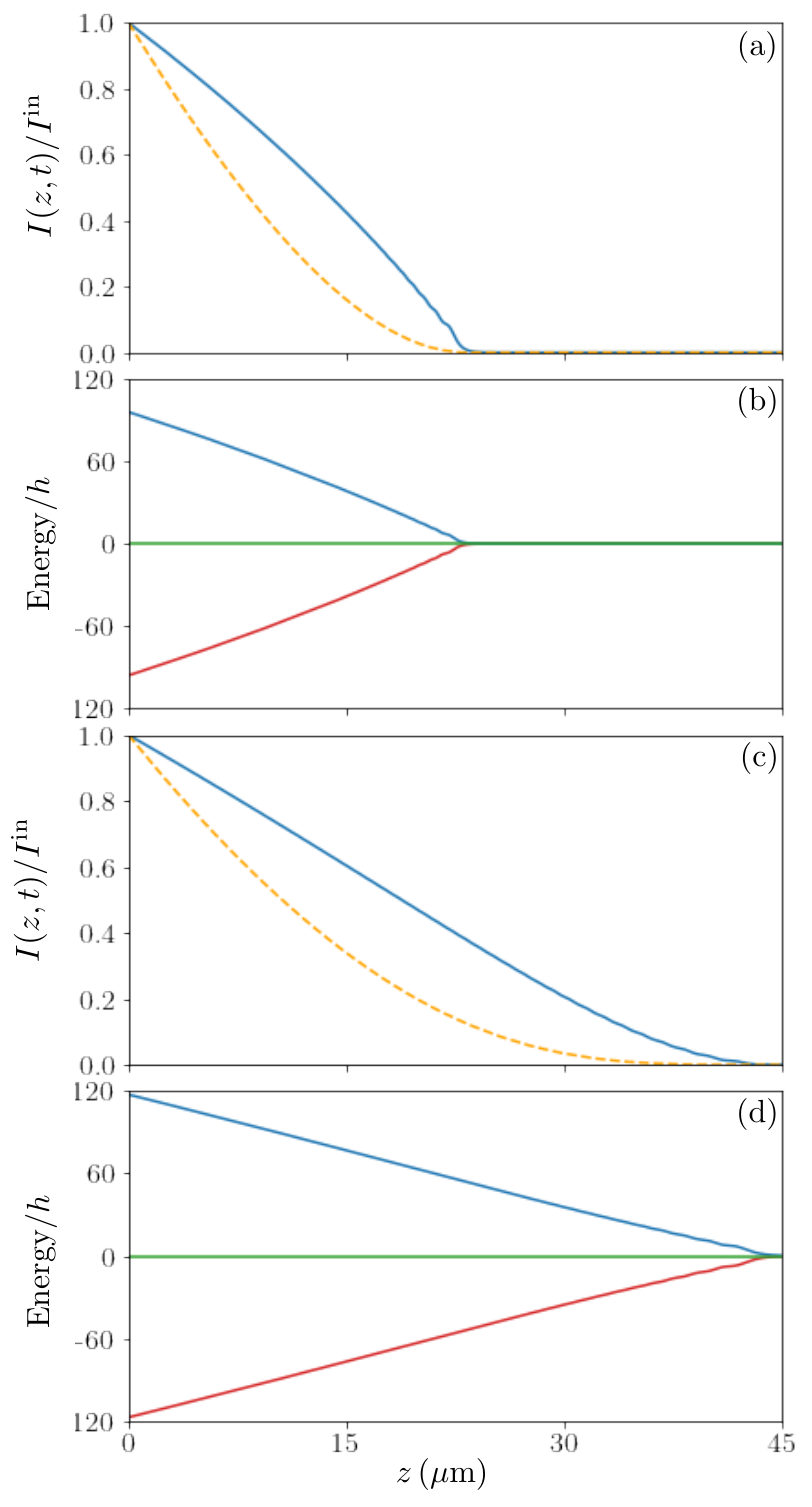}
    \caption{The fields and the eigenenergies of the rotating-wave Hamiltonian in the steady state regime. (a): The intensity of the probe field (solid curve) and of the coupling field (dashed curve) for $I_{\rm p}^{\rm in} = 1$~$\mu$W~cm$^{-2}$ and $I_{\rm c}^{\rm in} = 5$~W~cm$^{-2}$ at $t = 180$~ns. (b): The three eigenenergies of the rotating-wave Hamiltonian for the same fields as in (a). (c) and (d): The same as (a) and (b) but for $I_{\rm p}^{\rm in} = I_{\rm c}^{\rm in} = 5$~W~cm$^{-2}$.}
\label{fig:6}
\end{figure}

The intensities of these two fields in the steady-state regime are plotted as functions of $z$ in Fig.~\ref{fig:6}(a), for the same parameters as in Fig.~\ref{fig:4}. These results are calculated for the largest time available from the simulation, $t = 180$~ns. The steady-state regime is not yet reached for $z > 20$~mm at that time, which explains why the intensity of the probe field does not vary monotonically in this region. The intensity of the coupling field does not exponentially decrease, except in the large-$z$ region, because $I_{\rm c}$ much exceeds the saturation intensity of the respective transition for $z < 20$~mm: if not for Doppler broadening, $I_{\rm c}$ would decrease linearly  \cite{Cartwright2022,Siddons2014}.

The Hamiltonian of this 3-state system is given by Eq.~(\ref{eq:appH}) of the appendix. Its eigenenergies are plotted in part (b) of Fig.~\ref{fig:6} for the $z$-dependent values of $\Omega_{\rm p}$ and $\Omega_{\rm c}$ corresponding to the intensities plotted in part (a). The detunings $\Delta_{\rm p}$ and $\Delta_{\rm c}$ are both taken to be zero in these calculations.  It can be seen that these energies are split by almost 200~MHz at $z = 0$ (the entrance of the medium), which far exceeds the 5.750~MHz natural linewidth of the D$_1$ transition. The probe field is thus far off-resonance at $z = 0$.
These two eigenenergies both approach zero when $z$ increases, as the coupling field is absorbed and $\Omega_{\rm c}$ decreases, which results in an increasingly fast absorption of the probe field.

The corresponding results for the case of a strong probe field are shown in parts (c) and (d) of Fig.~\ref{fig:6}. The parameters of the system are the same as in Fig.~\ref{fig:5} here. Both fields contribute significantly to the AC Stark shift now, which leads to a still larger gap between the eigenenergies at the entrance of the medium, to a slower reduction of this Autler-Townes splitting when $z$ increases, and therefore to a longer distance of propagation in the medium. Each field thus induces a transparency on the other field. The energy eigenstates of this system can be thought as being doubly dressed in this situation \cite{doublydressed}.

\begin{figure}[t]
    \centering
    \includegraphics[width=0.48\textwidth]{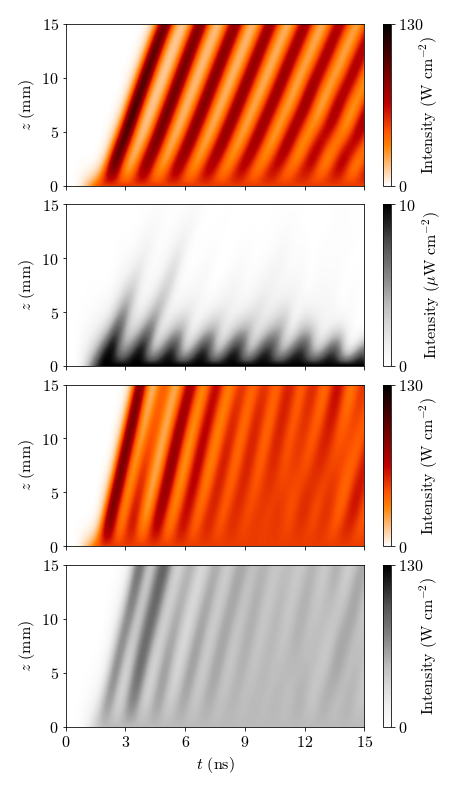}
    \caption{The coupling field and the probe field as obtained when the hyperfine structure of the 5S$_{1/2}$, 5P$_{1/2}$ and 5P$_{3/2}$ states is taken into account. The two fields co-propagate. (a) the coupling field and (b) and the probe field for $I_{\rm c}^{\rm in} = 50$~W~cm$^{-2}$ and $I_{\rm p}^{\rm in} = 10$~$\mu$W~cm$^{-2}$.
    (c) the coupling field and (d) the probe field for $I_{\rm c}^{\rm in} = I_{\rm p}^{\rm in} = 50$~W~cm$^{-2}$.}
    \label{fig:7}
\end{figure}
Co-propagating transients also occur in more complex V-systems, as illustrated by Fig.~\ref{fig:7}. The medium is a $^{85}$Rb vapor of the same density and Doppler temperature as in the above calculations; however, the system now comprises 46 states as the full hyperfine structure of the 5S$_{1/2}$, 5P$_{1/2}$ and 5P$_{3/2}$ states is now taken into account. The probe and coupling fields are resonant on, respectively, the 5S$_{1/2}(F=3)$~--~5P$_{1/2}(F=3)$ and
5S$_{1/2}(F=3)$~--~5P$_{3/2}(F=4)$ transitions.
The calculation follows Ref.~\cite{Ogden2019}, however here with the two fields being step pulses turned on simultaneously as prescribed by Eqs.~(\ref{eq:Eoft}) and (\ref{eq:Foft}). As seen from the figure, the incident fields spawn trains of transient pulses similar to those found in the 3-level model. However, the probe field does not follow the coupling field as closely. The difference is particularly marked when the applied probe field is weak [Fig.~\ref{fig:7}(a,b)], although even in this case the probe field still propagates significantly farther inside the medium than predicted by the Beer-Lambert law \cite{explainalpha}.
The two fields also co-propagate in step when both are sufficiently strong, although one can note a time- and position-dependent variation of the relative intensity of each pulse of the train [Fig.~\ref{fig:7}(c,d)].
    

\section{Conclusions}
\label{section:conclusions}

In summary, we have studied the transient pulse trains triggered by the sudden turn-on of a strong CW field applied to an atomic vapor. We have extended previous works on the formation of these transients both by examining the role of homogeneous broadening and by addressing their occurrence in more complex V-systems.

For the case of 2-level media with no homogeneous broadening, our results depart from the conclusions of previous investigations \cite{Crisp1972,Horovitz1982} in showing that the average time separation between the pulses increases with distance. However, we have confirmed that the pulse trains found in this system are well described by dnoidal solutions of the Maxwell-Bloch equation over a sufficiently short time interval. Incorporating homogeneous broadening into the model does not entirely suppress the formation of a dnoidal-like oscillation, which is consistent with experiment \cite{Segard1990,deLamare1993} and has been observed in previous calculations \cite{Macke2010}. However, each pulse then slows down and broadens as it propagates. The ensuing oscillation of the field is damped and transient, and disappears as the system evolves into its normal steady state.

Similarly, we have shown that switching on a doubly resonant CW field may create transient pulse trains akin to a twin pair of dnoidal waves. Soliton-induced transparency in V-systems generalizes into a ``dnoidal-wave-induced transparency" whereby a strong field  carries a weak field of a different frequency as a co-propagating dnoidal wave. An analytical model of this effect is outlined in the appendix for the case where all broadening mechanisms are negligible. However, our numerical results show that doubly resonant fields may propagate in step even in the presence of homogeneous broadening and Doppler broadening. Stable analytical dnoidal twin waves are also exact solutions of the Maxwell-Bloch equations in the case where the two fields are both strong, although then only under the conditions set by Eqs.~(\ref{eq:Delta}) and (\ref{eq:equal_mus}); however, our numerical results also show that similar transients may form for two strong fields even when neither of these conditions is fulfilled. The mode of transmission of the incident light changes from SIT to V-EIT as the system evolves into a steady state and the initial transients disappear. As shown by the results of Fig.~\ref{fig:7}, these effects can also be found in more complicated multi-state systems than those usually considered in this context.

Depending on the experimental conditions, representing the applied field by a plane wave, as done here, may not be appropriate for predicting the results of actual measurements. E.g., the duration and period of the transients may significantly vary across the width of the applied beam due to its transverse inhomogeneity \cite{deLamare1993,Macke2010}. The transient pulses described above may also undertake significant transverse reshaping due to diffraction and/or self-focusing \cite{transverse}. Nonetheless, these effects do not preclude the existence of the SIT transients considered in the present work, even if they may obscure their manifestation.



\acknowledgments

The authors thank L.\ A.\ Bol'shov for providing a copy of Ref.~\cite{preprint} and I.\ G.\ Hughes for insightful information about EIT in V-systems. BSC was supported by EPSRC under grant EP/T518001/1 during this work.


\begin{center}
{\small \bf DATA AVAILABILITY}
\end{center}
The data that support the findings of this article are openly available \cite{data}.


\appendix 

\section{Dnoidal waves in V-systems}
\label{appendix:Vdnoidal}
\subsection{Two-color fields}

This appendix extends Sec.~\ref{section:twostate} to the case considered in Sec.~\ref{section:V}, namely that of two fields co-propagating in a same direction and addressing a 3-level system in a V configuration. As usual in this context, we refer to one of these fields as the probe field and to the other as the coupling field. The former and the latter couple the ground state (state~1) to two different excited states (respectively state~2 and state~3). We neglect homogeneous broadening throughout this appendix. We therefore assume that these two excited states are stable over the time scale considered in this work.

We start by deriving the equations of motion for this particular case.
We describe each of the two fields by a real electric field vector, namely ${\bf E}_{\rm p}(z,t)$ for the probe field and ${\bf E}_{\rm c}(z,t)$ for the coupling field, the total electric field at time $t$ and distance $z$ from the front of the medium as being ${\bf E}(z,t)$ with
\begin{equation}
    {\bf E}(z,t) = 
{\bf E}_{\rm p}(z,t) +
{\bf E}_{\rm c}(z,t).
\label{eq:appEtot}
\end{equation}
We assume that these two fields are both linearly polarized and set
\begin{align}
{\bf E}_{\rm p}({z},t) &= \frac{1}{2}\, \hat{\text{\boldmath{$\epsilon$}}}_{\rm p}\, {\cal E}_{\rm p}(z,t)
\exp[-i\omega_{\rm p}(t - z/c)] + \mbox{c.c.},\\
{\bf E}_{\rm c}({z},t) &= \frac{1}{2}\, \hat{\text{\boldmath{$\epsilon$}}}_{\rm c}\, {\cal E}_{\rm c}(z,t)
\exp[-i\omega_{\rm c}(t - z/c)] + \mbox{c.c.},
\end{align}
with real unit polarization vectors $\hat{\text{\boldmath{$\epsilon$}}}_{\rm p}$ and $\hat{\text{\boldmath{$\epsilon$}}}_{\rm c}$.
Correspondingly, we make the following ansatz for the medium polarization,
\begin{align}
{\bf P}({z},t) = & \frac{1}{2}\,\hat{\text{\boldmath{$\epsilon$}}}_{\rm p} \, {\cal P}_{\rm p}({z},t)
\exp[-i\omega_{\rm p}(t - z/c)] + \nonumber \\
& \frac{1}{2}\,\hat{\text{\boldmath{$\epsilon$}}}_{\rm c} \, {\cal P}_{\rm c}({z},t)
\exp[-i\omega_{\rm c}(t - z/c)]
 + \mbox{c.c.}
\end{align}
Eq.~(\ref{eq:propap}) then generalizes to a pair of wave equations,
\begin{align}
\frac{\partial {\cal E}_{\rm p}}{\partial z} + \frac{1}{c}
\frac{\partial {\cal E}_{\rm p}}{\partial t} &= \frac{i\omega_{\alpha}}{2\epsilon_0 c}\,{\cal P}_{\rm p}(z,t), \label{eq:apppropap}\\
\frac{\partial {\cal E}_{\rm c}}{\partial z} + \frac{1}{c}
\frac{\partial {\cal E}_{\rm c}}{\partial t} &= \frac{i\omega_{\alpha}}{2\epsilon_0 c}\,{\cal P}_{\rm c}(z,t).\label{eq:apppropac}
\end{align}
These two equations are coupled to each other through the dependence of ${\cal P}_{\rm p}({z},t)$ and ${\cal P}_{\rm c}({z},t)$ on both ${\cal E}_{\rm p}(z,t)$ and ${\cal E}_{\rm c}(z,t)$.

As in Sec.~\ref{section:theory}, we treat the coupling of the atoms with the fields within the electric dipole approximation, make the rotating wave approximation and pass to slowly varying variables. This results in the Hamiltonian
\begin{align}
&\hat{H}({z},t) = - \hbar \Delta_{\rm p}|2 \rangle \langle 2 | - \hbar \Delta_{\rm c} | 3 \rangle \langle 3| \nonumber \\
 &\; -(\hbar/2) \big[\Omega_{\rm p}(z,t)
|2\rangle\langle 1| + \Omega_{\rm c}(z,t) |3 \rangle\langle 1| + \mbox{h.c.}\,\big],
\label{eq:appH}
\end{align}
with
\begin{equation}
    \Omega_{\rm p}(z,t) =
{\cal E}_{\rm p}(z,t)\,\langle\, 2\, |\,
\hat{\text{\boldmath{$\epsilon$}}}_{\rm p} \cdot \hat{\bf D} \,|\, 1\rangle/\hbar
\label{eq:appOmegap}
\end{equation}
and
\begin{equation}
    \Omega_{\rm c}(z,t) =
{\cal E}_{\rm c}(z,t)\,\langle\, 3\, |\,
\hat{\text{\boldmath{$\epsilon$}}}_{\rm c} \cdot \hat{\bf D} \,|\, 1\rangle/\hbar,
\label{eq:appOmegac}
\end{equation}
where $\hat{\bf D}$ is the dipole operator. We assume that the fields are on resonance in the absence of any Doppler shift. The detunings $\Delta_{\rm p}$ and $\Delta_{\rm c}$ may nonetheless differ from zero if Doppler broadening is included in the calculation: for atoms with a velocity $u_z$ in the $z$-direction, 
\begin{equation}
    \Delta_{\rm p} = -\omega_{\rm p}u_z/c, \qquad \Delta_{\rm c} = -\omega_{\rm c}u_z/c.
\end{equation}
In terms of the coherences $\rho_{21}(z,t)$ and  $\rho_{31}(z,t)$, which are respectively equal to
$\langle 2 | \hat{\rho}(z,t) | 1 \rangle$
and $\langle 3 | \hat{\rho}(z,t) | 1 \rangle$, where $\hat{\rho}(z,t)$ is the density operator for this 3-state system,
\begin{align}
    {\cal P}_{\rm p}(z,t) &= 2 N_{\rm d} \, 
    \langle\, 1\, |\,
\hat{\text{\boldmath{$\epsilon$}}}_{\rm p} \cdot \hat{\bf D} \,|\, 2\, \rangle\, \rho_{21}(z,t),
\label{eq:apppolp}\\
{\cal P}_{\rm c}(z,t) &= 2 N_{\rm d} \, 
    \langle\, 1\, |\,
\hat{\text{\boldmath{$\epsilon$}}}_{\rm c} \cdot \hat{\bf D} \,|\, 3\, \rangle\, \rho_{31}(z,t).
\label{eq:apppolc}
\end{align}
As in Sec.~\ref{section:theory}, $N_d$ denotes the medium number density. 
Combining these last two equations with Eqs.~(\ref{eq:apppropap}), (\ref{eq:apppropac}), (\ref{eq:appOmegap}) and (\ref{eq:appOmegac}) gives
\begin{align}
  \frac{\partial \Omega_{\rm p}}{\partial z} + \frac{1}{c}\,\frac{\partial \Omega_{\rm p}}{\partial t} &= i \mu_{\rm p} \,
    \rho_{21}(z,t) \label{eq:appOmegappropa}\\
 \frac{\partial \Omega_{\rm c}}{\partial z} + \frac{1}{c}\,\frac{\partial \Omega_{\rm c}}{\partial t} &= i \mu_{\rm c} \,
    \rho_{31}(z,t), \label{eq:appOmegacpropa}
\end{align}
where $\mu_{\rm p}$ and $\mu_{\rm c}$ are the respective propagation coefficients:
\begin{align}
    \mu_{\rm p} &= \frac{\omega_{\rm p}N_{\rm d}}{\hbar \epsilon_0 c}\, |\langle 2 |\,
\hat{\text{\boldmath{$\epsilon$}}}_{\rm p} \cdot \hat{\bf D} \,|\, 1\, \rangle|^2, \\
 \mu_{\rm c} &= \frac{\omega_{\rm c}N_{\rm d}}{\hbar \epsilon_0 c}\, |\langle 3 |\,
\hat{\text{\boldmath{$\epsilon$}}}_{\rm p} \cdot \hat{\bf D} \,|\, 1\, \rangle|^2.
\end{align}

It has long been recognized that these equations can be solved analytically when $\mu_{\rm p} = \mu_{\rm c}$, in the absence of relaxation and as long as inhomogeneous broadening is negligible or can be taken to be described by a same distribution function $g(\Delta)$ for both fields. (In the case of Doppler broadening, the latter is possible only if their frequencies are sufficiently close.) The solutions relevant to this paper are derived in the rest of this appendix.

\subsection{The Maxwell-Bloch equations without homogeneous broadening}
As noted in Sec.~\ref{section:theory}, calculating the density operator $\hat{\rho}(z,t)$ involves calculating the velocity-specific density operators $\hat{\rho}_v(x,t,u_z)$ for all the relevant velocity classes. 
Since we neglect homogeneous broadening here, we set
\begin{equation}
\hat{\rho}_v(z,t,u_z) \equiv |\Psi(z,t)\rangle\langle \Psi(z,t)|
\end{equation}
with
\begin{equation}
    |\Psi(z,t)\rangle = c_1(z,t)|1\rangle + c_2(z,t) |2\rangle + c_3(z,t) |3 \rangle,
\end{equation}
where the coefficients $c_1(z,t)$, $c_2(z,t)$ and $c_3(z,t)$ are complex functions such that
\begin{equation}
    |c_1(z,t)|^2 + |c_2(z,t)|^2 + |c_3(z,t)|^2 = 1.
\end{equation}
Like $\hat{\rho}_v({\bf r},t,u_z)$,
the state vector $|\Psi(z,t)\rangle$ and the functions $c_1(z,t)$, $c_2(z,t)$ and $c_3(z,t)$ depend on $u_z$ as well as on $z$ and $t$.
Given Eq.~(\ref{eq:appH}), these coefficients and the two fields evolve in time according to the equations
\begin{subequations}
\label{eq:allB3}
\begin{align}
i& \frac{\partial c_1}{\partial t} = -\frac{1}{2}\, \Omega_{\rm p}^*(z,t) c_2(z,t)  -\frac{1}{2}\, \Omega_{\rm c}^*(z,t) c_3(z,t), \label{eq:B3a}\\
i& \frac{\partial c_2}{\partial t} = -\Delta_{\rm p} c_2(z,t) -\frac{1}{2}\, \Omega_{\rm p}(z,t) c_1(z,t), \label{eq:B3b}\\
i& \frac{\partial c_3}{\partial t} = -\Delta_{\rm c} c_3(z,t) -\frac{1}{2}\, \Omega_{\rm c}(z,t) c_1(z,t), \label{eq:B3c}\\
&    \frac{\partial \Omega_{\rm p}}{\partial z} + \frac{1}{c}\,\frac{\partial \Omega_{\rm p}}{\partial t} = 
\nonumber \\ & \qquad \qquad
i \mu_{\rm p} \, \int_{-\infty}^{\infty} f_{}(u_z)\,
    c_2(z,t)c_1^*(z,t)\, {\rm d}u_z, \label{eq:B3d}\\
&   \frac{\partial \Omega_{\rm c}}{\partial z} + \frac{1}{c}\,\frac{\partial \Omega_{\rm c}}{\partial t} =
\nonumber \\ & \qquad \qquad
i \mu_{\rm c} \, \int_{-\infty}^{\infty} f_{}(u_z)\,
    c_3(z,t)c_1^*(z,t)\, {\rm d}u_z. \label{eq:B3e}
\end{align}
\end{subequations}
Eqs.~(\ref{eq:allB3}a)--(\ref{eq:allB3}e) do not have an analytical solution when $\Delta_{\rm p} \not= \Delta_{\rm c}$. As is commonly done, we will assume from now on that
\begin{equation}
    \Delta_{\rm p} = \Delta_{\rm c} = \Delta.
    \label{eq:Delta}
\end{equation}
This condition amounts to taking $\omega_{\rm p} = \omega_{\rm c} = \omega$ and is fulfilled only approximately for the systems for which numerical results are presented in this article. We note, however, that this condition is unnecessary if inhomogeneous broadening is neglected, in which case the equations (\ref{eq:allB3}a)--(\ref{eq:allB3}e) reduce to the following system:
\begin{subequations}
\label{eq:allB6}
\begin{align}
i& \frac{\partial c_1}{\partial t} = -\frac{1}{2}\, \Omega_{\rm p}^*(z,t) c_2(z,t)  -\frac{1}{2}\, \Omega_{\rm c}^*(z,t) c_3(z,t), \\
i& \frac{\partial c_2}{\partial t} = -\frac{1}{2}\, \Omega_{\rm p}(z,t) c_1(z,t), \\
i& \frac{\partial c_3}{\partial t} = -\frac{1}{2}\, \Omega_{\rm c}(z,t) c_1(z,t), \\
&  \frac{\partial \Omega_{\rm p}}{\partial z} + \frac{1}{c}\,\frac{\partial \Omega_{\rm p}}{\partial t} = i \mu_{\rm p} \,
    c_2(z,t)c_1^*(z,t)\\
&  \frac{\partial \Omega_{\rm c}}{\partial z} + \frac{1}{c}\,\frac{\partial \Omega_{\rm c}}{\partial t} = i \mu_{\rm c} \,
    c_3(z,t)c_1^*(z,t).
\end{align}
\end{subequations}

It follows from Eqs.~(\ref{eq:allB3}a)--(\ref{eq:allB3}e) and (\ref{eq:Delta}) that
\begin{subequations}
\label{eq:conservation_nobr}
\begin{align}
\frac{\partial |\Omega_{\rm p}|^2}{\partial z}
+\frac{1}{c}\,
\frac{\partial |\Omega_{\rm p}|^2}{\partial t}
    &= -2 \mu_{\rm p} \int_{-\infty}^\infty \frac{\partial |c_2|^2}{\partial t}\,g(\Delta)\,{\rm d}\Delta,\\
    \frac{\partial |\Omega_{\rm c}|^2}{\partial z}
+\frac{1}{c}\,
\frac{\partial |\Omega_{\rm c}|^2}{\partial t}
    &= -2 \mu_{\rm c} \int_{-\infty}^\infty \frac{\partial |c_3|^2}{\partial t}\,g(\Delta)\,{\rm d}\Delta,
\end{align}
\end{subequations}
where we treat the coefficients $c_2(z,t)$ and $c_3(z,t)$ as functions of $\Delta$ rather than $u_z$ and, as in Sec.~\ref{section:theory},
\begin{align}
    g_{}(\Delta) & \equiv (c/\omega)f_{}(-c \Delta / \omega).
\end{align}

We consider solutions which depend on $z$ and $t$ only through the dimensionless variable $(t - z/v)/\tau$, where $v$ and $\tau$ are two constants. For such solutions,
\begin{subequations}
\label{eq:conservation_eta}
\begin{align}
    \left(  
    \frac{1}{v} - \frac{1}{c}\right)
    \frac{\partial |\Omega_{\rm p}|^2}{\partial t}
    &= 2 \mu_{\rm p} \int_{-\infty}^\infty \frac{\partial |c_2|^2}{\partial t}\,g(\Delta)\,{\rm d}\Delta,\\
    \left(  
    \frac{1}{v} - \frac{1}{c}\right)
    \frac{\partial |\Omega_{\rm c}|^2}{\partial t}
    &= 2 \mu_{\rm c} \int_{-\infty}^\infty \frac{\partial |c_3|^2}{\partial t}\,g(\Delta)\,{\rm d}\Delta.
\end{align}
\end{subequations}
In the case of soliton-like solutions, for which $\Omega_{\rm p}(z,t)$, $\Omega_{\rm c}(z,t)$, $c_2(z,t)$ and $c_3(z,t)$ vanish for $t \rightarrow \pm \infty$, but not in the case of dnoidal wave solutions, these last two equations imply that
\begin{subequations}
\label{eq:conservation_integrated}
\begin{align}
    \left(  
    \frac{1}{v} - \frac{1}{c}\right)
    |\Omega_{\rm p}(z,t)|^2
    &= 2 \mu_{\rm p} \int_{-\infty}^\infty |c_2(z,t)|^2\,g(\Delta)\,{\rm d}\Delta,\\
    \left(  
    \frac{1}{v} - \frac{1}{c}\right)
    |\Omega_{\rm c}(z,t)|^2
    &= 2 \mu_{\rm c} \int_{-\infty}^\infty |c_3(z,t)|^2\,g(\Delta)\,{\rm d}\Delta.
\end{align}
\end{subequations}
We also note that when inhomogeneous broadening is neglected,
Eqs.~(\ref{eq:conservation_nobr}a)--(\ref{eq:conservation_nobr}b) reduce to
\begin{subequations}
\begin{align}
\frac{\partial |\Omega_{\rm p}|^2}{\partial z}
+\frac{1}{c}\,
\frac{\partial |\Omega_{\rm p}|^2}{\partial t}
    &= -2 \mu_{\rm p}  \frac{\partial |c_2|^2}{\partial t},\\
    \frac{\partial |\Omega_{\rm c}|^2}{\partial z}
+\frac{1}{c}\,
\frac{\partial |\Omega_{\rm c}|^2}{\partial t}
    &= -2 \mu_{\rm c} \frac{\partial |c_3|^2}{\partial t},
\end{align}
\end{subequations}
Eqs.~(\ref{eq:conservation_eta}a)--(\ref{eq:conservation_eta}b) reduce to
\begin{subequations}
\begin{align}
    \left(  
    \frac{1}{v} - \frac{1}{c}\right)
    \frac{\partial |\Omega_{\rm p}|^2}{\partial t}
    &= 2 \mu_{\rm p} \frac{\partial |c_2|^2}{\partial t},\\
    \left(  
    \frac{1}{v} - \frac{1}{c}\right)
    \frac{\partial |\Omega_{\rm c}|^2}{\partial t}
    &= 2 \mu_{\rm c} \frac{\partial |c_3|^2}{\partial t},
\end{align}
\end{subequations}
and Eqs.~(\ref{eq:conservation_integrated}a)--(\ref{eq:conservation_integrated}b) reduce to \cite{Bolshov1982b}
\begin{subequations}
\begin{align}
    \left(  
    \frac{1}{v} - \frac{1}{c}\right)
    |\Omega_{\rm p}(z,t)|^2
    &= 2 \mu_{\rm p} |c_2(z,t)|^2,\\
    \left(  
    \frac{1}{v} - \frac{1}{c}\right)
    |\Omega_{\rm c}(z,t)|^2
    &= 2 \mu_{\rm c} |c_3(z,t)|^2.
\end{align}
\end{subequations}

\subsection{Equal propagation coefficients}

The above equations can be solved in closed form when the propagation coefficients $\mu_{\rm p}$ and $\mu_{\rm c}$ are equal. In particular, setting
\begin{equation}
\mu_{\rm p} = \mu_{\rm c} = \mu.
\label{eq:equal_mus}
\end{equation}
makes it possible to seek solutions of either Eqs.~(\ref{eq:allB3}a)--(\ref{eq:allB3}e) or Eqs.~(\ref{eq:allB6}a)--(\ref{eq:allB6}e) for which the two fields are in a constant ratio, i.e., solutions for which
\begin{equation}
    \Omega_{\rm p}(z,t) \equiv r \Omega_{\rm c}(z,t),
\end{equation}
where $r$ is a real constant. For instance, setting $c_2(z,t) \equiv r c_3(z,t)$ and passing to the dependent variables $a(z,t)$, $b(z,t)$ and $\Omega(z,t)$ such that $c_1(z,t) \equiv a(z,t)$, $c_3(z,t) \equiv b(z,t)/(1 + r^2)^{1/2}$ and
$\Omega_{\rm c}(z,t) \equiv \Omega(z,t)/(1 + r^2)^{1/2}$ reduces Eqs.~(\ref{eq:allB6}a)--(\ref{eq:allB6}e) to the following system, which describes the propagation of a single field in a 2-level medium \cite{Rahman}:
    \begin{subequations}
\label{eq:all2stnoinh}
\begin{align}
i& \frac{\partial a}{\partial t} = -\frac{1}{2}\, \Omega_{}^*(z,t) b(z,t),\\
i& \frac{\partial b}{\partial t} = -\frac{1}{2}\, \Omega_{}(z,t) a(z,t),\\
&    \frac{\partial \Omega_{}}{\partial z} + \frac{1}{c}\,\frac{\partial \Omega_{}}{\partial t} =
i \mu\,
    b(z,t)a^*(z,t).
    \end{align}
\end{subequations}
Taking $a(z,t)$, $b(z,t)$ and $\Omega(z,t)$ to correspond to the sech-soliton solution of this last system of equations yields Konopnicki and Eberly's sech-simulton solution for a V-system \cite{Rahman}. Other choices are also possible. In particular, the dnoidal solution is obtained by taking
\begin{subequations}
\label{eq:allB11}
\begin{align}
    a(z,t) &= -\mbox{sn}(\eta,k)\\
    b(z,t) &= i\,\mbox{cn}(\eta,k),\\
    \Omega(z,t) &= \Omega_0 \, \mbox{dn}(\eta,k),
\end{align}
\end{subequations}
where
\begin{equation}
    \eta = (t-z/v)/\tau
\end{equation}
with
\begin{equation}
\frac{1}{v} = \frac{1}{c} + \frac{\mu\tau^2}{2 k^2}
\end{equation}
and
\begin{equation}
\label{eq:taunew}
    \tau = 2 / \Omega_0.
\end{equation}
The corresponding Rabi frequencies for the probe and the coupling fields 
are, respectively,
$\Omega_{{\rm p}0} \, \mbox{dn}(\eta,k)$ and $\Omega_{{\rm c}0} \, \mbox{dn}(\eta,k)$ with
\begin{equation}
\sqrt{\Omega^2_{{\rm p}0} + \Omega^2_{{\rm c}0}} = \Omega_0.
\end{equation}
This solution corresponds to the twin pair of co-propagating dnoidal waves described by Eqs.~(\ref{eq:vdnoidal}), (\ref{eq:25}), (\ref{eq:26}) and (\ref{eq:tau2colour}). It is a particular case of a still more general solution found for V-systems in the absence of any broadening mechanism \cite{Hioe1994}. Taking $k \rightarrow 1$ yields
the simulton solution of
Konopniki and Eberly \cite{Kono81}, namely ${\Omega}_{\rm p}(z,t) = 
    {\Omega}_{{\rm p}0}\,\mbox{sech}(\eta)$
and ${\Omega}_{\rm c}(z,t) = 
    {\Omega}_{{\rm c}0}\,\mbox{sech}(\eta)$.

Similarly, expressing Eqs.~(\ref{eq:allB3}a)--(\ref{eq:allB3}e) in terms of the coefficients $a(z,t)$ and $b(z,t)$ and of the Rabi frequency $\Omega(z,t)$ reduces these equations to the system 
\begin{subequations}
\label{eq:all2st}
\begin{align}
i& \frac{\partial a}{\partial t} = -\frac{1}{2}\, \Omega_{}^*(z,t) b(z,t),\\
i& \frac{\partial b}{\partial t} = -\Delta_{} b(z,t) -\frac{1}{2}\, \Omega_{}(z,t) a(z,t),\\
&    \frac{\partial \Omega_{}}{\partial z} + \frac{1}{c}\,\frac{\partial \Omega_{}}{\partial t} = 
\nonumber \\ & \qquad \qquad
i \mu \, \int_{-\infty}^{\infty} g{}(\Delta)
    b(z,t)a^*(z,t)\, {\rm d}\Delta.
    \end{align}
\end{subequations}
The dnoidal solution is now given by the following equations, where $X(\eta)$, $Y(\eta)$ and $Z(\eta)$ are the functions defined by Eqs.~(8b)--(8d) of Ref.~\cite{Crisp1969}:
\begin{subequations}
\label{eq:allB11withinhbr}
\begin{align}
    a(z,t) &= \sqrt{[1 - Z(\eta)]/2}\; \exp[i\alpha(\eta)],\\
    b(z,t) &= 
    \sqrt{[1+Z(\eta)]/2}\; \exp[i\beta(\eta)],\\
    \Omega(z,t) &= \Omega_0 \, \mbox{dn}(\eta,k),
\end{align}
\end{subequations}
with \cite{noteaboutCrisp1969}
\begin{align}
    \alpha(\eta) &= \pi + \int_0^\eta \frac{X(\eta')\,\mbox{dn}(\eta',k)}{1-Z(\eta')}\,{\rm d}\eta'
\end{align}
and
\begin{align}
\beta(\eta) &= \alpha(\eta) - \arctan[Y(\eta)/X(\eta)] + \pi.
\end{align}
The time parameter $\tau$ is still given by Eq.~(\ref{eq:taunew}), and $\Omega(z,t)$ is still given by Eq.~(\ref{eq:allB11}c), but $v$ is now defined by
Eq.~(\ref{eq:vdnoidalib}) of Sec.~\ref{section:theory}. 

The coefficients $c_1(z,t)$, $c_2(z,t)$ and $c_3(z,t)$ and the electric field amplitudes ${\cal E}_{\rm p}(z,t)$ and ${\cal E}_{\rm c}(z,t)$ so defined depend on $z$ and $t$ only through the variable $\eta$. These five functions are periodic in $\eta$ with period $2K(k)$. They are thus periodic in $t$ with period $2\tau K(k)$ and in $z$ with period $2v\tau K(k)$, and the coefficients $c_1(z,t)$, $c_2(z,t)$ and $c_3(z,t)$ oscillate in phase across all the velocity classes. The two fields also satisfy the area relation
\begin{align}
&\sqrt{
    \left[\int_{-\tau K(k)}^{\tau K(k)} \Omega_{\rm p}(z,t)\, {\rm d}t\right]^2 + 
    \left[\int_{-\tau K(k)}^{\tau K(k)} \Omega_{\rm c}(z,t)\, {\rm d}t\right]^2 
    }  = 2\pi,
\end{align}
which generalizes a similar relation applying only to single sech-simultons, i.e.,
\begin{align}
&\sqrt{
    \left[\int_{-\infty}^{\infty} \Omega_{\rm p}(z,t)\, {\rm d}t\right]^2 + 
    \left[\int_{-\infty}^{\infty} \Omega_{\rm c}(z,t)\, {\rm d}t\right]^2 
    }  = 2\pi.
\end{align}

This dnoidal wave solution does not exhaust the possibilities. For instance, a cnoidal wave solution would be obtained by setting
\begin{equation}
    \Omega(z,t) = \Omega_0\,\mbox{cn}(\eta,k)
\end{equation}
and taking the functions $X(\eta)$, $Y(\eta)$ and $Z(\eta)$ appearing in the above equations to be those defined by Eqs.~(16b)--(16d) of Ref.~\cite{Crisp1969}. In the absence of inhomogeneous broadening, V-systems also support the propagation of pairs of fields for which both $\Omega_p(z,t)$ and $\Omega_c(z,t)$ are linear combinations of a cn function and a dn function but are not in a constant ratio \cite{Hioe1994}. Solutions of this latter type have no counterpart in two-level systems and might not exist for Doppler-broadened systems.

\subsection{Unequal propagation coefficients}
\label{appendix:unequal}

Whether inhomogeneous broadening is taken into account or not, $\Omega_{\rm p}(z,t)$ and $\Omega_{\rm c}(z,t)$ cannot be obtained in closed form without further approximation if the propagation coefficients $\mu_{\rm p}$ and $\mu_{\rm c}$ are unequal.
Exact results applicable to the case where $\mu_{\rm p} \not= \mu_{\rm c}$ are scarce. To the best of our knowledge, they are limited to the following relations, which pertain to the soliton-like solutions of Eqs.~(\ref{eq:allB6}a)--(\ref{eq:allB6}e) \cite{Bolshov1982b}:
\begin{subequations}
\label{eq:Bolshov}
\begin{align}
    \Omega_{\rm c}(z,t) &= \Omega_{{\rm c}0} \exp[\xi(\eta)], \\
   \Omega_{\rm p}(z,t) &= \Omega_{{\rm p}0} \exp[\gamma\xi(\eta)],  
\end{align}
\end{subequations}
where $\xi(\eta)$ is a certain function of $\eta$ (not known in closed form) which goes to zero for $\eta \rightarrow \pm \infty$ and
\begin{equation}
\gamma = (\mu_{\rm p}/\mu_{\rm c})^{1/2}.
\label{eq:gammadefined}
\end{equation}
This system can thus support twin pulses co-propagating indefinitely even when the two propagation coefficients are different, with the probe pulse being longer (shorter) than the coupling pulse if $\gamma < 1$ ($ > 1$). Unfortunately, these results are of limited applicability here, as their derivation assumes that $\Omega_{\rm p}(z,t)$, $\Omega_{\rm c}(z,t)$ and the coefficients $c_j(z,t)$ depend on $z$ and $t$ only through the variable $\eta$ defined above, with constant values of $v$ and $\tau$, and that $\Omega_{\rm p}(z,t)$,  $\Omega_{\rm c}(z,t)$, $c_2(z,t)$ and $c_3(z,t)$ vanish for $t \rightarrow -\infty$. The first assumption excludes the transients studied in this work, for which the effective pulse velocity is not constant. The second excludes stationary dnoidal waves. Also, it seems unlikely that these results can be generalized to the inclusion of inhomogeneous broadening.

However, solving the Maxwell-Bloch equations analytically for a V-system with $\mu_{\rm p} \not= \mu_{\rm c}$ is possible for zero inhomogeneous broadening if one assumes that the probe field is too weak to affect the propagation of the coupling field. To the best of our knowledge, this was first done by Kozlov, Polynkin, and Scully, for a pair of fields co-propagating as a sech-simulton \cite{Kozlov1999}. In the following, we generalize their calculation to the case where both fields co-propagate as dnoidal waves, although working in terms of wave functions rather than populations and coherences. The approach followed here may not be applicable to cases where inhomogeneous broadening needs to be taken into account.

\begin{figure}[h!]
    \centering
    \includegraphics[width=0.48\textwidth]{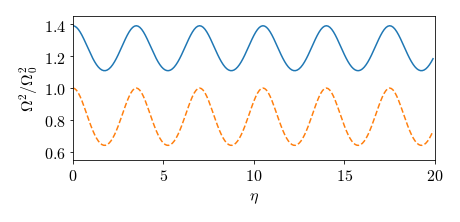}
    \caption{The ratio $\Omega^2_{\rm p}/\Omega^2_{{\rm p}0}$ (solid curve) and the ratio $\Omega^2_{\rm c}/\Omega^2_{{\rm c}0}$ (dashed curve) for $k = 0.6$ and $\gamma = 0.7$ as calculated from Eqs.~(\ref{eq:Omegacappendix}) and (\ref{eq:Omegapweak3}).}
    \label{fig:comparison}
\end{figure}

Besides ignoring all broadening mechanisms, the key approximation is to replace Eq.~(\ref{eq:allB6}a) by
\begin{subequations}
\label{eq:newB3a}
\begin{equation}
    i \frac{\partial c_1}{\partial t} =  -\frac{1}{2}\, \Omega_{\rm c}^*(z,t) c_3(z,t).
\end{equation}
\end{subequations}
Eqs.~(\ref{eq:allB6}c), (\ref{eq:allB6}e) and (\ref{eq:newB3a}a) can then be solved separately from Eqs.~(\ref{eq:allB6}b) and
(\ref{eq:allB6}d), to obtain the dnoidal solution
\begin{subequations}
\label{eq:couplingfield}
\begin{align}
    c_1(z,t) &= -\mbox{sn}(\eta,k)\\
    c_3(z,t) &= i\,\mbox{cn}(\eta,k),\\
    \Omega_{\rm c}(z,t) &= \Omega_{{\rm c}0} \, \mbox{dn}(\eta,k), \label{eq:Omegacappendix}
\end{align}
\end{subequations}
here with $\tau$ and $v$ defined by the equations
\begin{equation}
    \tau = 2/\Omega_{{\rm c}0}
\end{equation}
and
\begin{equation}
\frac{1}{v} = \frac{1}{c} + \frac{\mu_{\rm c}\tau^2}{2 k^2}.
\label{eq:newv}
\end{equation}
In turn, we seek a solution of Eqs.~(\ref{eq:allB6}b) and
(\ref{eq:allB6}d) which depends on $z$ and $t$ only through the variable $\eta$. Combining these two equations gives
\begin{align}
    &\left(\frac{1}{c} - \frac{1}{v}\right)\, \frac{1}{\tau}\,
    \frac{{\rm d} \Omega_{\rm p}}{{\rm d} \eta} = \nonumber \\
    &\quad- \frac{1}{2}\,\mu_{\rm p}\,c_1^*(\eta) \tau \left[ \int_{\eta_0}^\eta \Omega_{\rm p}(\eta')c_1(\eta')\,{\rm d}\eta' + C\,\right],
\end{align}
where $\eta_0$ is a constant and $C = -2i c_2(\eta_0)/\tau$.
That is, in view of Eq.(\ref{eq:newv}) and the fact that $c_1(\eta)$ is real,
\begin{equation}
\frac{{\rm d}\Omega_{\rm p}}{{\rm d}\eta} =
\gamma^2 k^2 c_1(\eta) \left[\int_{\eta_0}^\eta \Omega_{\rm p}(\eta')c_1(\eta')\,{\rm d}\eta' + C\,\right],
\end{equation}
or, after rearrangement, differentiation and division by $c_1(\eta)$,
\begin{equation}
\frac{1}{c_1(\eta)}\,\frac{{\rm d} \;}{{\rm d} \eta} \frac{1}{c_1(\eta)}
    \frac{{\rm d}\Omega_{\rm p}}{{\rm d}\eta} = \gamma^2 k^2 \Omega_{\rm p}(\eta).
    \label{eq:transformed}
\end{equation}
At this stage, we introduce a function $x(\eta)$ such that
\begin{equation}
    \frac{{\rm d}x}{{\rm d}\eta} = c_1(\eta),
\label{eq:B42}
\end{equation}
and write $\Omega_{\rm p}$ as a function of $x$ rather than a function of $\eta$. Given that $c_1(\eta) = -\mbox{sn}(\eta,k)$ here, 
\begin{equation}
    x(\eta) = \frac{1}{k}\,
    \left\{ \log \left[\frac{\mbox{dn}(\eta,k)+k\,\mbox{cn}(\eta,k)}{k'}\right]\right\} + C',
\end{equation}
where $C'$ is a constant and
\begin{equation}
k' = \sqrt{1-k^2}.    
\end{equation}
It can be shown that the function $\mbox{dn}(\eta,k)+k\,\mbox{cn}(\eta,k)$ is positive 
for any $\eta$ when $0 \leq k < 1$ and varies monotonically on the interval $0 \leq \eta \leq 2K(k)$. We therefore
exclude the limiting case of $k=1$ from our analysis and, for the time being, restrict ourselves to seeking a solution of Eq.~(\ref{eq:transformed}) on that interval of values of $\eta$. Rewriting this equation in terms of the variable $x$ yields
\begin{equation}
    \frac{{\rm d}^2\Omega_{\rm p}}{{\rm d}^2x} = 
    \gamma^2 k^2 \Omega_{\rm p}(x).
\end{equation}
Thus
\begin{align}
    \Omega_{\rm p}(\eta) = 
    &\; C_+ \left[\frac{\mbox{dn}(\eta,k)+k\,\mbox{cn}(\eta,k)}{k'} \right]^{\gamma} + \nonumber \\ 
    &\; C_- \left[\frac{\mbox{dn}(\eta,k)+k\,\mbox{cn}(\eta,k)}{k'} \right]^{-\gamma},
\label{eq:Omegapweak1}
\end{align}
where $C_+$ and $C_-$ are two constants.
Eq.~(\ref{eq:Omegapweak1}) can also be written as
\begin{align}
    \Omega_{\rm p}(\eta) = 
    & \; C_{+} \left[\frac{\mbox{dn}(\eta,k)+k\,\mbox{cn}(\eta,k)}{k'} \right]^{\gamma} + \nonumber \\
    & \; C_{-} \left[\frac{\mbox{dn}(\eta,k)-k\,\mbox{cn}(\eta,k)}{k'} \right]^{\gamma}
\label{eq:Omegapweak2}
\end{align}
in view of the identity
\begin{equation}
    \frac{k'}{\mbox{dn}(\eta,k)+k\,\mbox{cn}(\eta,k)} =
    \frac{\mbox{dn}(\eta,k)-k\,\mbox{cn}(\eta,k)}{k'}.
\end{equation}
Since $\mbox{dn}(\eta,k)$ is periodic in $\eta$ with period $2K(k)$ whereas $\mbox{cn}[\eta+2K(k),k] = -\mbox{cn}(\eta,k)$, continuing $\Omega_{\rm p}(\eta)$ as a periodic function of $\eta$ with the same period as $\Omega_{\rm c}(\eta)$ requires that $C_+=C_-$. Thus
\begin{align}
    \Omega_{\rm p}(z,t) &= \Omega_{{\rm p}0}\left\{\left[{\mbox{dn}(\eta,k)+k\,\mbox{cn}(\eta,k)} \right]^{\gamma} \right. \nonumber \\
    &\qquad \qquad \left. + \left[{\mbox{dn}(\eta,k)-k\,\mbox{cn}(\eta,k)} \right]^{\gamma}\right\}/2^\gamma
\label{eq:Omegapweak3}
\end{align}
with $\Omega_{{\rm p}0} = (2/k')^\gamma C_+$. It then follows from Eqs.~(\ref{eq:allB6}b) and (\ref{eq:B42}) that
\begin{align}
    c_2(z,t) &= i (\Omega_{{\rm p}0}/\Omega_{{\rm c}0})\left\{\left[{\mbox{dn}(\eta,k)+k\,\mbox{cn}(\eta,k)} \right]^{\gamma} \right. \nonumber \\
    &\qquad \left. - \left[{\mbox{dn}(\eta,k)-k\,\mbox{cn}(\eta,k)} \right]^{\gamma}\right\}/(2^\gamma \gamma k).
\end{align}
These results are consistent with those obtained above for $\gamma=1$, bearing in mind that we assume that $\Omega_{{\rm p}0} \ll \Omega_{{\rm c}0}$ here.
As illustrated by Fig.~\ref{fig:comparison}, $\Omega_{\rm p}(z,t)$ oscillates with the same period and the same phase as $\Omega_{\rm c}(z,t)$ even when $\gamma \not= 1$.
We also note that
\begin{equation}
    \Omega_{\rm p}(z,t) \rightarrow \Omega_{{\rm p}0}\,[\mbox{sech}(\eta)]^\gamma
\end{equation}
in the limit $k \rightarrow 1$, in agreement both with Eq.~(36) of Ref.~\cite{Kozlov1999} and with Eqs.~(\ref{eq:Bolshov}a) and
(\ref{eq:Bolshov}b) above.


\begin{thebibliography}{99}
%
\bibitem{McCall}
S.\ L.\ McCall and E.\ L.\ Hahn,
Phys.\ Rev.\ Lett.\ {\bf 18}, 908 (1967); Phys.\ Rev.\ {\bf 183}, 457 (1969).
%
\bibitem{Crisp1969}
M. D. Crisp, Phys. Rev. Lett. {\bf 22}, 820 (1969).
%
\bibitem{Eberly1969}
J. H. Eberly, Phys. Rev. Lett. {\bf 22}, 760 (1969).
%
\bibitem{Lamb1971}
G. L. Lamb, Jr., Rev. Mod. Phys. {\bf 43}, 99 (1971).
%
\bibitem{Bolshov1985}
L.\ A.\ Bol'shov and V.\ V.\ Likhanski{\u \i}, Kvantovaya Elektron.\ {\bf 12}, 1339 (1985) [Sov.\ J.\ Quantum Electron.\ {\bf 15}, 889 (1985)].
%
\bibitem{Maimistov1990}
A.\ I.\ Maimistov, A.\ M.\ Basharov, S.\ O.\ Elyutin, and Yu.\ M.\ Sklyarov, Phys. Rep. {\bf 191}, 1 (1990); A.\ I.\ Maimistov, and A.\ M.\ Basharov, {\it Nonlinear Optical Waves},
Kluwer, Dordrecht (1999).
%
\bibitem{seeAllen1975}
For an introduction to the early work on SIT, see, e.g., L.\ Allen and J.\ H.\ Eberly,
{\it Optical Resonances and Two-level Atoms}, Wiley, New York (1975).
%
\bibitem{Ogden2019}
T.\ P.\ Ogden, K.\ A.\ Whittaker, J.\ Keaveney, S.\ A.\ Wrathmall, C.\ S.\ Adams, and R.\ M.\ Potvliege, Phys. Rev. Lett. {\bf 123}, 243604 (2019).
%
\bibitem{Arkhipov2020}
M.\ V.\ Arkhipov, A.\ A.\ Shimko, N.\ N.\ Rosanov, I. Babushkin, and R. M. Arkhipov, Phys. Rev. A {\bf 101}, 013803 (2020).
%
\bibitem{Bai2020}
Z.\ Bai, C.\ S.\ Adams, G.\ Huang, and W. Li, Phys. Rev. Lett. {\bf 125}, 263605 (2020).
%
\bibitem{othermedia}
See, e.g., C. Jirauschek, M. Riesch, and P. Tzenov, Adv. Theory Simul. {\bf 2}, 1900018 (2019);
G. T. Adamashvili, Eur. Phys. J. D {\bf 74}, 41 (2020); 
S. K. Hazra, P. K. Pathak, and T. N. Dey, Phys. Rev. B {\bf 107} 235409 (2023); 
M. S. Najafabadi, L. L. S{\'a}nchez-Soto, J. F. Corney, N. Kalinin, A. A. Sorokin, and G. Leuchs, Phys. Rev. Res. {\bf 6}, 023142 (2024).
%
\bibitem{femto}
See, e.g.,
%
S.\ Hughes, Phys. Rev. Lett. {\bf 81}, 3363 (1998); 
%
D.\ V.\ Novitsky, Phys. Rev. A {\bf 84}, 013817 (2011); 
%
H.\ Wu, J.\ Tang, M.\ Chen, M.\ Xiao, Y.\ Lu, K.\ Xia, and F. Nori, Opt. Express {\bf 32}, 11010 (2024); 
%
A.\ Pakhomov, Phys. Rev. A {\bf 111}, 013502 (2025).
%
\bibitem{Alhasan1992}
A.\ M.\ Alhasan, J.\ Fiutak, and W.\ Miklaszewski, Z. Phys. B {\bf 88}, 349 (1992).
%
\bibitem{Miklaszewski1994}
W.\ Miklaszewski and J.\ Fiutak, Z. Phys. B {\bf 93}, 491 (1994).
%
\bibitem{Kono81}
M.\ J.\ Konopnicki and J.\ H.\ Eberly,
Phys.\ Rev.\ A {\bf 24}, 2567 (1981).
%
\bibitem{seealso}
See also M.\ J.\ Konopnicki, P.\ H.\ Drummond, and
J.\ H.\ Eberly, Opt.\ Commun.\ {\bf 36}, 313 (1981);
C.\ R.\ Stroud, Jr.\ and D.\ A.\ Cardimona,
Opt.\ Commun.\ {\bf 37}, 221 (1981).
%
\bibitem{Huang}
Many-state V-systems may also support many-color simultons --- see, e.g.,
G. Huang and C. Hang, Phys. Lett. A {\bf 354}, 406 (2006);
G. Huang, C. Hang, and L. Deng, Eur. Phys. J. D {\bf 40}, 437 (2006).
%
\bibitem{Hioe1994}
F. T. Hioe and R. Grobe, Phys. Rev. Lett. {\bf 73}, 2559 (1994).
%
\bibitem{Crisp1972}
M. D. Crisp, Phys. Rev. A {\bf 5}, 1365 (1972).
%
\bibitem{Segard1990}
B. S{\'e}gard, B. Macke, J. Zemmouri, and W. Sergent, Ann. Phys.
(Paris) {\bf 15}, 167 (1990).
%
\bibitem{deLamare1993}
J. de Lamare, Ph. Kupecek, and M. Comte, Opt. Commun. {\bf 95}, 305 (1993).
%
\bibitem{Crisp1970}
M. D. Crisp, Phys. Rev. A {\bf 1}, 1604 (1970); ibid. {\bf 2}, 2172 (1970).
%
\bibitem{Rothenberg1984}
J. E. Rothenberg, D. Grischkowsky, and A. C. Balant,
Phys. Rev. Lett. {\bf 53}, 552 (1984).
%
\bibitem{Avenel1984}
O. Avenel, E. Varoquaux, and G. A. Williams, Phys. Rev. Lett. {\bf 53}, 2058 (1984).
%
\bibitem{Pessina1991}
E. M. Pessina, B. S{\'e}gard, and B. Macke, Opt. Comm. {\bf 81}, 397 (1991).
%
\bibitem{LeFew2009}
W. R. LeFew, S. Venakides, and D. J. Gauthier, Phys. Rev. A {\bf 79}, 063842 (2009).
%
\bibitem{Wei2009}
D. Wei, J. F. Chen, M. M. T. Loy, G. K. L. Wong, and S. Du, Phys. Rev. Lett. {\bf 103}, 093602 (2009).
%
\bibitem{Macke2010}
B. Macke and B. S{\'e}gard, Phys. Rev. A {\bf 81}, 015803 (2010).
%
\bibitem{precursorpapers}
K. E. Oughstun, N. A. Cartwright, D. J. Gauthier, and H. Jeong, J. Opt. Soc. Am. B {\bf 27}, 1664 (2010);
B. Macke and B. S{\'e}gard, {\it ibid.} {\bf 28}, 450 (2011); K. E. Oughstun, N. A. Cartwright, D. J. Gauthier, and H. Jeong, {\it ibid.} {\bf 28}, 468 (2011).
%
\bibitem{Horovitz1982}
B. Horovitz and N. Rosenberg, Phys. Rev. A {\bf 26}, 2799 (1982).
%
%
\bibitem{Kaup1977}
D. J. Kaup, Phys. Rev. A {\bf 16}, 704 (1977).
%
\bibitem{variapulses}
M. A. Newbold and G. J. Salamo, Phys. Rev. Lett. {\bf 42}, 887 (1979);
J. L. Shultz and G. J. Salamo, Phys. Rev. Lett. {\bf 78}, 855 (1997);
M.\ O.\ Scully, G.\ S.\ Agarwal, O.\ Kocharovskaya, V.\ V.\ Kozlov,
and A.\ B.\ Matsko, Opt.\ Express {\bf 8}, 66 (2001);
S. M. Saadeh, J. L. Shultz, and G. J. Salamo, Opt. Express {\bf 8}, 153 (2001).
%
\bibitem{Cartwright2022} B. S. Cartwright, Optical transients in atomic vapours, Durham theses, Durham University (2022), http://etheses.dur.ac.uk/14532/
%
\bibitem{Lambropoulos2007}
See, e.g., P. Lambropoulos and D. Petrosyan, {\it Fundamentals of Quantum Optics and Quantum Information}, Springer, Berlin (2007). No time-derivative of the polarization amplitude ${\cal P}(z,t)$ appears in Eq.~(\ref{eq:propap}) because the corresponding terms are negligible compared to the term in $\omega {\cal P}(z,t)$ forming the right-hand side of this equation. For the systems considered in the present paper, the amplitudes ${\cal E}(z,t)$ and ${\cal P}(z,t)$ vary on a time scale roughly equivalent to $10^6$ optical periods.
%
\bibitem{warning}
These results are readily derived by particularizing the calculations outlined in Appendix~\ref{appendix:Vdnoidal} to the case of a single field. They only apply to systems for which the function $g(\Delta)$ appearing in Eq.~(\ref{eq:vdnoidalib}) is an even function of $\Delta$. They need to be supplemented by a dispersion relation in more general cases \cite{McCall}.
%
\bibitem{Kdefined}
We follow the NIST Digital Library of Mathematical Functions (https://dlmf.nist.gov/) in the definition of this function:
\begin{displaymath}
    K(z) = \int_0^{\pi/2} \frac{{\rm d}\theta}{\sqrt{1-z^2\sin^2\theta}}.
\end{displaymath}
%
\bibitem{Kozlov2009}
V.\ V.\ Kozlov and E.\ B.\ Kozlova, Opt.\ Spektrosk.\ {\bf 107}, 139 (2009)
[Opt.\ Spectrosc.\ {\bf 107}, 129 (2009)].
%
\bibitem{Bolshov1982b}
L. A. Bol'shov, N. N. Elkin, T. K. Kirichenko, V. V. Likhanski{\u \i}, and A. P. Napartovich, Kvantovaya Elektron.\ {\bf 9},
1476 (1982) [Sov.\ J.\ Quantum Electron.\ {\bf 12}, 941 (1982)].
%
\bibitem{preprint}
L. A. Bol'shov, N. N. Elkin, T. K. Kirichenko, V. V. Likhanski{\u \i}, and M. I. Persiantsev, Preprint IAE-3732/16, Atomic Energy Inst., Moscow, 1983 [in Russian].
%
\bibitem{Bolshov1988}
L.\ A.\ Bol'shov, N.\ N.\ Yelkin, V.\ V.\ Likhanski{\u \i}, and
M.\ I.\ Persiantsev, Zh.\ Eks.\ Teor.\ Fiz.\ {\bf 94}, 101 (1988)
[Sov.\ Phys.\ JETP {\bf 67}, 2013 (1988)].
%
\bibitem{Kozlov1998}
V.\ V.\ Kozlov and E.\ E.\ Fradkin, Pis'ma Zh.\ Eksp.\ Teor.\ Fiz.\ {\bf 68},
359 (1998) [JETP Lett.\ {\bf 68}, 383 (1998)].
%
\bibitem{Denisova1998}
N.\ V.\ Denisova, V.\ S.\ Egorov, V.\ V.\ Kozlov, N.\ M.\ Reutova,
P.\ Yu.\ Serdobintsev, and E.\ E.\ Fradkin, Zh.\ Eksp.\ Teor.\ Fiz.\ {\bf 113}, 71
(1998) [J.\ Exp.\ Theor.\ Phys.\ {\bf 86}, 39 (1998)].
%
\bibitem{Kozlov1999}
V. V. Kozlov, P. G. Polynkin, and M. O. Scully, Phys. Rev. A {\bf 59}, 3060 (1999).
%
\bibitem{Paspalakis2000}
E. Paspalakis, N. J. Kylstra, and P. L. Knight, Phys. Rev. A {\bf 61}, 045802 (2000).
%
\bibitem{KK2010}
V.\ V.\ Kozlov and E.\ B.\ Kozlova, Opt.\ Spektrosk.\ {\bf 108}, 824 (2010)
[Opt.\ Spectrosc.\ {\bf 108}, 780 (2010)].
%
\bibitem{Fedotova2014}
O. M. Fedotova, O. K. Khasanov, G. A. Rusetsky, J. Degert, and E. Freysz, Phys. Rev. A {\bf 90}, 053843 (2014).
%
%
\bibitem{Potvliege2025}
R.\ M.\ Potvliege and S.\ A.\ Wrathmall, Comput. Phys. Commun. {\bf 306}, 109374 (2025).
%
\bibitem{Steck}
D.\ A. Steck, Rubidium 85 D line data, available online at the URL http://steck.us/alkalidata
%
%
\bibitem{Grieneisen1972}
H. P. Grieneisen, J. Goldhar, N. A. Kurnit, and A. Javan, Appl. Phys. Lett. {\bf 21}, 559 (1972).
%
\bibitem{Hamadani1974}
S. M. Hamadani, J. Goldhar, N. A. Kurnit, and A. Javan, Appl. Phys. Lett. {\bf 25}, 160 (1974).
%
\bibitem{Matusovsky1996b}
M. Matusovsky, B. Vaynberg, and M. Rosenbluh, J. Opt. Soc. Am. B {\bf 13}, 1994 (1996).
%
\bibitem{explainalpha}
According to the Beer-Lambert law, the  intensity of the probe field would decrease like $I_{\rm p}^{\rm in}\exp(-\alpha z)$ with $\alpha = (0.42\;\mbox{mm})^{-1}$ in the case of Fig.~\ref{fig:4}(b) and $\alpha=(1.2\;\mbox{mm})^{-1}$ in the case of Fig.~\ref{fig:7}, in the absence of the coupling field. See, e.g., Ref.~\cite{Potvliege2025} for the calculation of the absorption coefficient $\alpha$ with Doppler broadening.
%
\bibitem{Higgins2021}
E.g., 
C. R. Higgins and I. G. Hughes, J. Phys. B: At. Mol. Opt. Phys. {\bf 54}, 165403 (2021).
%
\bibitem{AbiSalloum2010}
T. Y. Abi-Salloum, Phys. Rev. A {\bf 81}, 053836 (2010).
%
\bibitem{Khan2016}
S. Khan, V. Bharti, and V. Natarajan, Phys. Lett. A {\bf 380}, 4100 (2016).
%
\bibitem{Zhu2013}
Quantum interference plays a role for weak coupling fields, though. See C. Zhu, C. Tan, and G. Huang, Phys. Rev. A {\bf 87}, 043813 (2013).
%
\bibitem{Siddons2014}
P. Siddons, J. Phys. B: At. Mol. Opt. Phys. {\bf 47}, 093001 (2014).
%
\bibitem{doublydressed}
Doubly dressed states have been considered previously for ladder systems [see, e.g., R.-Y. Chang, W.-C. Fang, Z.-S. He, B.-C. Ke, P.-N. Chen, and C.-C. Tsai, Phys. Rev. A {\bf 76}, 053420 (2007)] but, to our knowledge, not for V-systems.
%
\bibitem{transverse}
It is known that SIT solitons and simultons undergo transverse reshaping upon propagation: see
N. Wright and M. C. Newstein, Opt. Commun. {\bf 9}, 8 (1973);
H.\ M.\ Gibbs, B.\ B\"olger, F.\ P.\ Mattar, M.\ C.\ Newstein, G.\ Forster, and P.\ E.\ Toschek, Phys.\ Rev.\ Lett.\ {\bf 37}, 1743 (1976);
L.\ A.\ Bol'shov, V.\ V.\ Likhanski{\u{\i}}, and A.\ P.\ Napartovich, Zh.\ Eksp.\ Teor.\ Fiz.\ {\bf 72}, 1769 (1977) [Sov.\ Phys.\ JETP {\bf 45}, 928 (1977)];
F. P. Mattar and M. C. Newstein, Comput. Phys. Comm. {\bf 20}, 139 (1980);
L.\ A.\ Bol'shov, T.\ K.\ Kirichenko, V.\ V.\ Likhanski{\u{\i}}, M.\ I.\ Persiantsev, and L.\ K.\ Sokolova, Zh.\ Eksp.\ Teor.\ Fiz.\ {\bf 86}, 1240 (1984) [Sov.\ Phys.\ JETP {\bf 59}, 724 (1984)];
P.\ D.\ Drummond, Opt.\ Commun.\ {\bf 49}, 219 (1984); J. de Lamare, M. Comte, and Ph. Kupecek, Phys. Rev. A {\bf 50}, 3366 (1994); J. de Lamare, Ph. Kupecek, and M. Comte, Phys. Rev. A {\bf 51}, 4289 (1995).
%
\bibitem{data}
The propagated fields and the corresponding input data for use with the open-source Maxwell-Bloch solver CoOMBE \cite{Potvliege2025} can be found at the URL https://doi.org/10.15128/r2jq085k062.
%
\bibitem{Rahman}
A.\ Rahman, Phys. Rev.\ A {\bf 60}, 4187 (1999);
A.\ Rahman and J.\ H.\ Eberly, Opt.\ Express {\bf 4}, 133 (1999).
%
\bibitem{noteaboutCrisp1969}
These equations take into the fact that the symbol $\Delta$ used in Ref.~\cite{Crisp1969} denotes the negative of the detuning $\Delta$ defined in the present work.
%

%
%
%
%
%
%
%
%
%
%
%
%
%
%
%
%
%
%
%
%
%
%

%
%
%
%
\end{thebibliography}
\end{document}